\documentclass[aps,pra,reprint,floatfix,superscriptaddress,showkeys,amsmath,amssymb,longbibliography]{revtex4-1}
\usepackage[english]{babel}
\usepackage{amsmath}
\usepackage{amssymb}
\usepackage{natbib}
\usepackage{graphicx}
\usepackage{bm}

\usepackage{subfigure}
\usepackage{float}
\usepackage{epsfig}
\usepackage{footnote}
\usepackage{hyperref}
\hypersetup{
colorlinks=true,final=true,
        linkcolor=blue,
        citecolor=blue,
        filecolor=blue,
        urlcolor=blue,
}

\def\be{\begin{equation}}
\def\ee{\end{equation}}
\def\ba{\begin{eqnarray}}
\def\ea{\end{eqnarray}}

\newcommand{\btext}[1]{\textcolor{black}{{#1}}}

\begin{document}

\title{Topological Magnus responses in two and three dimensional systems}

\date{\today}

\author{Sanjib Kumar Das$^\clubsuit$}\email{s.k.das@ifw-dresden.de}
\affiliation{IFW Dresden and W{\"u}rzburg-Dresden Cluster of Excellence ct.qmat, Helmholtzstr. 20, 01069 Dresden, Germany}
\author{Tanay Nag$^\clubsuit$} \email{tnag@physik.rwth-aachen.de}
\affiliation{Institute f\"ur Theorie der Statistischen Physik, RWTH Aachen University, 52056 Aachen, Germany}
\author{Snehasish Nandy}\email{sn5jm@virginia.edu}
\affiliation{Department of Physics, University of Virginia, Charlottesville, VA 22904 USA}
\thanks{$^\clubsuit$Both authors SKD and TN contributed equally.}

\begin{abstract}
 
Recently, time-reversal symmetric but inversion broken systems with non-trivial Berry curvature in the presence of a built-in electric field have been proposed to exhibit a new type of linear Hall effect in ballistic regime, namely, the Magnus Hall effect. The transverse current here is caused by the Magnus velocity that is proportional to the built-in
electric field enabling us to examine the 
Magnus responses, in particular, Magnus Hall conductivity and Magnus Nernst conductivity, with chemical potential.
 Starting with  two-dimensional (2D) topological systems, we find that 
warping induced asymmetry in both the Fermi surface and Berry curvature can in general enhance the Magnus response for monolayer graphene and surface states of topological insulator. The strain alone is only responsible for  Magnus valley responses in monolayer graphene while warping leads to finite Magnus response there.
Interestingly, on the other hand, strain can change the Fermi surface character substantially that further results in distinct behavior of  Magnus transport coefficients as we observe in bilayer graphene. These responses there remain almost insensitive to warping unlike  the case of monolayer graphene.
Going beyond 2D systems, we also investigate the Magnus  responses in three-dimensional multi-Weyl semimetals (mWSMs) to probe the effect of tilt and anisotropic nonlinear energy dispersion. 
Remarkably, Magnus responses can only survive for the WSMs with  chiral
tilt. In particular, our study indicates that the  chiral (achiral) tilt engenders  Magnus (Magnus valley) responses.
Therefore, Magnus responses can be used as a tool  to distinguish between the untilted and tilted WSMs in experiments. 
In addition, we find that the Magnus Hall responses get suppressed with increasing  the nonlinearity associated with the  band touching around multi-Weyl node. 
\end{abstract}

\maketitle

\section{Introduction}
\label{intro}

The family of Hall effects have revolutionized  the solid state physics in the context of novel electronic states and electron dynamics \cite{hall79,karplus54,klitzing80}. Starting from the Lorentz force mediated classical Hall effect and quantum Hall effect, various new types of Hall effect such as anomalous Hall effect, spin Hall effect, thermal Hall effect etc. have been discovered over the years \cite{hall79,karplus54,klitzing80,haldane88,Sinitsyn_2007,
nagaosa10,Chang167,liu16,he18,kane05,Sinova_2015}. Among them, the Berry curvature (BC) induced Hall effects, taking place without external magnetic field, have drawn tremendous attention to both the theorists and experimentalists \cite{haldane88,Sinitsyn_2007,nagaosa10,Chang167,liu16,
he18,kane05,Niu_2010}.
For time reversal symmetry (TRS) broken systems, the BC takes the form $\bm \Omega(\bm k)\ne -\bm \Omega (-\bm k)$ leading to a finite total BC for the occupied states. By contrast,  in TRS invariant systems,  BC follows $\bm \Omega(\bm k)= -\bm \Omega (-\bm k)$ giving rise to zero total BC \cite{Niu_2010}, that further causes the intrinsic anomalous Hall effect to vanish. Interestingly, it has been theoretically proposed and
experimentally verified that unlike the linear anomalous Hall effect, the nonlinear anomalous Hall effect can survive in TRS invariant systems with broken inversion symmetry \cite{sodemann15,Xu2018,Ma2018,Facio18,Yu19,Du2018,Du2019,Du2021,Ortix2021}. Very recently, the same systems (TRS invariant but inversion broken) with a built-in
electric field at zero magnetic field is found to exhibit a
new type of a linear response namely, Magnus Hall effect (MHE) \cite{papaj19}. The MHE is
originated from the Magnus velocity of electron
that is perpendicular to the BC and the built-in electric field. 
Along with the prediction of MHE, it has also been proposed that the Magnus Nernst effect (MNE) and Magnus thermal Hall effect can  also appear in these systems in the presence of applied thermal gradiant~\cite{Mandal20}.

It has been proposed that the two-dimensional (2D) transition metal dichalcogenides MX$_2$ (M=Mo, W and X=S, Se, Te) \cite{Qian1344,Xiao12,You18,Xu2018}, monolayer (ML) graphene on hBN, bilayer (BL) graphene with applied perpendicular electric field \cite{McCann_2013,ROZHKOV2016,Battilomo19},
heterostructures \cite{Yankowitz2012} and surfaces of topological insulator (TI) \cite{Fu09} are the possible candidates to investigate the MHE and MNE for their TRS invariant and inversion broken nature. The MHE has been theoretically studied recently in graphene and transition metal dichalcogenides using the low-energy model without warping (except bilayer graphene where trigonal warping is considered) and strain \cite{papaj19,Mandal20}. On the other hand, it has been shown that without spin-orbit coupling and tilting of Dirac cone, the BC dipole becomes substantially large in presence of strain and warping for graphene \cite{Battilomo19}.  Motivated by the BC dipole induced nonlinear anomalous Hall effect in TRS invariant systems, our interest here is to investigate Magnus transport in the presence of strain and warping for the above proposed  suitable  candidates.

Turning to the field of three-dimensional (3D) topological systems such as,
Weyl semimetals (WSMs), considered to be a 3D analogue of graphene, have been 
studied extensively for their intriguing properties and anomalous response functions. The gap closing points, guaranteed by some crystalline symmetries, in WSMs
are referred to as Weyl nodes with topological charge $n=1$
\cite{Ran_11,Leon_11, Murakami_2007,Murakami_n,Burkov11,Wan11,Armitage18}. 
There exists two Weyl nodes of opposite chiralities for the TRS broken WSM, while inversion broken WSM exhibits at least four Weyl nodes \cite{Zyuzin12,McCormick17}. 
The WSMs can also be classified as type-I and type-II. In the case of type-I WSM, Fermi surface is always point like irrespective of the tilting of the node. On the other hand, in type-II WSM finite electron and hole pockets appear at the Fermi level as
a result of finite tilting of the energy spectra\cite{VOLOVIK2014514,YXu15,Soluyanov2015}.
Moreover, it has been recently found that $n>1$ multi-WSM (mWSM) shows non-linear band touching \cite{Gxu11,CFang12}.
The WSMs are shown to exhibit many intriguing transport properties, originated by chiral anomaly, such as negative longitudinal magnetoresistance and planar Hall effect \cite{Kim14,HOSUR2013,Dong_2015,VA_2017,Son_2013,
Burkov_2017,Burkov_2015,Wang2016,Zhang2016,
Nandy17,Sun_2018,Mandal_2018,Ghosh_2019, Kumar_2018}. Tilting of energy dispersion and non-linearity of band touching further decorate the transport signatures \cite{Udagawa16,Fei17,Dantas2018,Nag_2020,Nag20}. This motivates us to extend our investigation of MHE to 3D topological systems considering a generic  mWSM Hamiltonian. 

In this work, 
we first capture  intriguing Fermi surface phenomena in presence of strain and warping by examining  MHE and MNE in ballistic
regime for 2D topological systems such as,  ML, BL graphene and surface states of TIs. We find that in strained ML graphene without warping, the total valley integrated  Magnus responses are zero as the
contribution coming from individual valley exactly cancels each other. The valley polarized contribution thus leads to the Magnus valley Hall, Magnus valley Nernst effects. Interestingly, 
warping induces  valley integrated  finite  Magnus responses
as the asymmetries in  Fermi surface and BC  result in unequal valley polarized contributions. The magnitude of the Magnus responses enhance
with increasing the warping parameter.
The same is observed 
for the surface states of TI in presence of
 hexagonal warping.
On the other hand, for BL graphene, the Magnus transport coefficients are substantially  modified depending on the  positive and negative values of strain while warping do not affect the Magnus transport.      
Finally, going beyond 2D systems, we study Magnus responses in 3D WSMs to examine the effect of tilt and anisotropic nonlinear dispersion. 
We  find that the MHE is identically zero for each Weyl node without tilt. Remarkably, chiral (achiral) tilt causes finite 
MH and MN conductivities to generate from individual Weyl nodes resulting in   Magnus (Magnus Valley) responses.
Moreover, our study indicates that 
the topological charge associated with Weyl node imprints its effect on the Magnus transport properties.

The rest of the paper is organized as follows. In Sec.~\ref{formalism},
we derive the general expressions of MH and MN  conductivities in both ballistic and diffusive regimes. Following which, in sec.~\ref{results_MHE}
we have calculated the Magnus transport responses in the presence of strain and warping (tilt and non-linearity) for different 2D (3D) topological systems, respectively. Finally, we summarize our results and discuss possible
future directions in Sec.~\ref{cons}.

\section{Formalism of Magnus transport}
\label{formalism}

In this section, we derive the general expression for MH, MN  conductivities in both diffusive and ballistic regimes using Boltzmann transport equation. To begin with, we consider mesoscopic systems of electronic transport in a Hall bar device without applying any external magnetic field. 
In this setup, the source and the drain are kept at different  electrostatic potential energy with the gate voltages given by $U_s$ and $U_d$, respectively.  Their difference $\Delta U= U_s-U_d$ introduces a built-in electric field ${\bm E}_{\rm in}={\bm \nabla}_\mathbf{r} U/e$ ($-e$ is the electronic charge) in the device with a slowly varying electric potential energy $U(r)$  along the length of the sample.

Now in the presence of external electric field $\mathbf{E}$ and
temperature gradient $\mathbf{\nabla T}$ applied between the source and drain, the charge current $\mathbf{J}$ and thermal current $\mathbf{Q}$ from linear response theory, can be written as
\begin{eqnarray}
&J_a=\sigma_{ab}E_b + \alpha_{ab} (-\nabla_b T) \nonumber \\
&Q_a = \bar \alpha_{ab} E_b + \kappa_{ab}(-\nabla_b T),
\label{lrt}
\end{eqnarray}
where $a$ and $b$ are spatial indices running over $x$, $y$, $z$. Here ${\sigma}$, ${\alpha}$ and ${\kappa}$ different conductivity tensors. 

The phenomenological Boltzmann transport equation can be written as~\cite{ashcroft1976solid,ziman2001electrons}
\begin{equation}
\left(\frac{\partial}{\partial t}+\mathbf{\dot{r}}\cdot\mathbf{\nabla_{r}}+{\dot{\bm k}}
\cdot {\nabla_{\bm k}}\right)f_{{\bm k},\mathbf{r},t}=I_{coll} \{f_{{\bm k},\mathbf{r},t}\},
\label{eq_BZ}
\end{equation}
where the right side $I_{coll} \{f_{{\bm k},\mathbf{r},t}\}$ is the collision integral which
incorporates the effects of electron correlations and impurity scattering. The electron distribution function is denoted by  $f_{\mathbf{k},\mathbf{r},t}$. Now under the relaxation time
approximation the steady-state Boltzmann equation can be written as
\begin{equation}
(\mathbf{\dot{r}}\cdot\mathbf{\nabla_{r}}+{\dot{\bm k}}\cdot{\nabla_{\bm k}})f_{{\bm k}}=\frac{f_{0}-f_{{\bm k}}}{\tau({\bm k})},
\label{eq_BZf}
\end{equation}
where $\tau(\bm k)$ is the scattering time. Note that in this work, we ignore the  momentum dependence of $\tau(\bm k)$ for simplifying the calculations and assume it to be a constant. The equilibrium distribution function $f_0$ in absence of applied electric field ${\bm E}$ and temperature gradient ${\bm \nabla}_{r} T$ is given by the Fermi function,  
\be 
f_0({\bm k},{\bm r}) = \dfrac{1}{1 + e^{\beta\left[\epsilon({\bm k},{\bm r}) - \mu\right]} }~,
\label{fd_statics}
\ee
where $\beta = 1/(k_B T)$, $\epsilon({\bm k},{\bm r}) = \epsilon_ {\bm k} + U({\bm r})$, with $\epsilon_ {\bm k}$ and ${\mu}$ are the energy dispersion and chemical potential, respectively. The motion of the wave packet inside the Hall bar is described by the semiclassical equations of motion  \cite{Xiao2010,Son12}
\ba 
\hbar \dot {\bm r} &=& \nabla_{\bm k}\epsilon_k
+\left[\nabla_{\bm r}{U} + e{\bm E}\right] \times {\bm \Omega}~ ,\label{eq_rdot}
\\
\hbar \dot {\bm k} &=&-\nabla_{\bm r}{U} - e {\bm E}~.
\label{eq_kdot}
\ea
The first, second and third term in the right hand side of Eq.~ (\ref{eq_rdot})
respectively represent the band velocity, Magnus velocity $V_{\rm magnus}=\nabla_{\bm r}{U} \times \bm \Omega $ and anomalous velocity $V_{\rm ano}= {\bm E} \times {\bm \Omega} $. The Magnus velocity $V_{\rm magnus}$ can be  thought of a quantum analog of the classical Magnus effect.

Now to calculate MH and MN conductivities, we apply the electric field  and temperature gradient along $x$ direction. Assuming the length of the sample is along $x$ axis, we have $U(\mathbf{r})=U(x)$, $\epsilon({\bm k},{\bm r}) = \epsilon_ {\bm k} + U(x)$ and ${\bm E}_{\rm in}=\frac{1}{e} \frac{\partial U(x)}{\partial x}\hat{x}$. Since we consider $U(x)$ is slowly varying, the electron wave
packets traveling inside the sample still have well-defined momentum ${\bm k}$. 
Considering the velocity of an incident electron $(v_x,v_y)$, transit time through the electric
field region becomes $t=L/v_x$ where $L$ is device length along $x$-direction. For $v_y \ne 0$,
the center
of the wave packet receives a transverse shift (in $y$-direction), followed by  Magnus velocity proportional to ${\nabla}_x U  \Omega_z $,
while traversing the junction due to the built-in electric field ${\bm E}_{\rm in}$.
Now the charge and thermal currents can be written as
\be 
\left\{{\bf J}({x}) , {\bf Q}({x})\right\}=  \int d{\bm k}~ \dot{\bm r} \left\{-e, [\epsilon({\bm k}, {x}) - \mu] \right \}f({\bm k}, {x}) ~.
\label{current}
\ee

Combining Eqs.~(\ref{eq_BZf}), (\ref{eq_rdot}), and (\ref{eq_kdot}) the non-equilibrium distribution function $f$ up to linear order in the bias fields can be written as
\be 
f = f_0 + v_x \tau \big(eE_x +  \frac{\left[ \epsilon({\bm k},{\bm r} )-\mu\right]}{T} \nabla_x T\big) \partial_\epsilon f_0~.
\label{NDF_tau}
\ee
 Considering $U$ as a slowly varying function of $x$, $\partial U/\partial x = \Delta U/L$, one can obtain $f$ to be spatially independent. To be precise, using 
 Eq.~(\ref{NDF_tau}) into the Eq.~(\ref{current}) and comparing with Eq.~(\ref{lrt}),
the MH conductivity $\sigma$ is found  to be \cite{Mandal20}
\be 
\sigma = - \dfrac{e^2 \tau}{\hbar} \dfrac{\Delta U}{L} \int d{\bm k} ~\Omega_z v_{x} ~ \partial_\epsilon f_{0},
\label{mgnus_hll}
\ee
Similarly, the MN conductivity $\alpha$ is given by  \cite{Mandal20}
\begin{eqnarray}
&&\alpha =   \dfrac{e k_B \tau}{\hbar} \dfrac{\Delta U}{L} \int d{\bm k}~\Omega_z v_{x} \beta ( \epsilon_k -\mu)~  \partial_\epsilon f_{0} 
\label{mgnus_thc}
\end{eqnarray}
where we have neglected the contributions coming from the band velocity. Interestingly, as discussed above these Magnus responses are dependent on the built-in electric field.

Considering the limit $\mu \gg \frac{1}{\beta}$ in Eq.~(\ref{mgnus_hll})-(\ref{mgnus_thc}), the Wiedemann-Franz law  and Mott relation allow us to compute 
 MN conductivity $\alpha$  and Magnus thermal Hall (MTH) conductivity $\kappa$ alternatively ~\cite{ashcroft1976solid,xiao06,dong20}
\be
\alpha = -\frac{\pi^2k_B^2T}{3e}\frac{\partial\sigma}{\partial\mu}~,~~~~{\rm and}~~~~\textcolor{black}{{\kappa} = \frac{\pi^2k_B^2T}{3e^2}\sigma}~.
\ee

Having discussed above the diffusive limit, we shall now illustrate the Magnus responses in the ballistic regime. Since in this regime, the mean free time between two collisions is infinite $\tau \rightarrow \infty$ i.e., essentially no collision occur in the transport direction inside the length $L$ along the $x$-direction of the Hall bar. Therefore, the right hand side of the Boltzmann transport equation given in Eq.~(\ref{eq_BZ}) vanishes in the ballistic regime. In this setup, the carriers from the source with only positive velocity $v_x >0$ are allowed in region $0<x<L$. Now the ansatz for the non-equilibrium  distribution function is the following
\be 
f({\bm k}, {\bm r}) =
\begin{cases}
-\Delta \mu \partial_\epsilon f_0 - \frac{\epsilon({\bm k}, {\bm r})-{\mu}}{ T}\Delta T \partial_\epsilon f_0~~~\text {for}~~~v_x > 0~,\\
~~~~~~~~~~~0 ~~~~~~~~~~~~~~~~~~~~~~~~~~~\text {for } ~~v_x < 0~.
\label{ndf_E}
\end{cases} 
\ee
Comparing the Eq.~(\ref{NDF_tau}) and Eq.~(\ref{ndf_E}), one can identify  the scattering length $v_x \tau$ with the device 
length $L$ so that $-eL E_x =\Delta \mu$ and $\Delta T/L=-\nabla_x T$. The MH and MN coefficients in ballistic regime can be obtained as \cite{papaj19,Mandal20}
\be
\sigma =- \frac{e^2}{\hbar} \Delta U \int_{v_x>0} d{\bm k}~ \Omega_z  \partial_\epsilon f_{0}~,
\label{sigma}
\ee
\be
\alpha = \frac{e}{\hbar T}  \Delta U \int_{v_x>0} d{\bm k} ~\Omega_z \left( \epsilon_k- \mu  \right)  \partial_\epsilon f_{0}~.
\label{alpha}
\ee
Similar to the diffusive regime, the Magnus transport coefficients in the ballistic regime also obey the Mott relation and the Wiedemann-Franz law.

We would now like to add a few comments on Magnus responses. It is clear from the Eqs.~(\ref{mgnus_hll})-(\ref{mgnus_thc})  as well as Eqs.~(\ref{sigma})-(\ref{alpha}) that the MHE and MNE  are purely determined by
Fermi surface properties as they incorporate $\partial_\epsilon f_{0}$ factor. In order to obtain finite Magnus responses, the systems should have asymmetric Fermi surface. In addition, the  system must possess finite BC.
Moreover, the condition $v_x >0$ happens to be very important while summing the BC over the Brillouin zone. Combining all these, 
the active momentum modes ${\bm k}_v$, over the Fermi surface $\epsilon_k=\mu$, for which $v_x>0$ would contribute to Magnus transport.
The remaining momentum modes ${\bm k}- {\bm k}_v$ become inert in this ballistic transport and BC for these momentum modes do not determine the transport. 
These transports thus allow us to scan different Fermi surfaces by tuning $\mu$ and get the idea about angular distribution of BC within a given Fermi surface.

The Magnus responses can be regarded as an effective second order transport as the built-in electric field $\Delta U$ and external bias both come into calculation of currents. However, in terms of
the external bias field, it is still a linear response phenomena. 
The BC dipole induced second order transport \cite{sodemann15}, given by \textcolor{black}{$D_{ij}=\int d\bm k ~d_{ij}~ \delta(\epsilon_{\bm k}-\mu )$ with $d_{ij}= v_i(\bm k) ~\Omega_j(\bm k) $}, is primarily different from  Magnus transport for which the tilting of Dirac cone is no longer important to obtain a finite response. 
\textcolor{black}{Depending upon the mirror symmetries present in the systems, one can find symmetry permitted components of non-linear transport coefficients where $d_{ij}$ becomes an even function of ${\bm k}$ \cite{zhang18,Yu19,zeng21}. By contrast, quantized  non-linear response namely, circular photo-galvanic effect can be observed for mirror symmetry broken non-centrosymmetric systems \cite{de2017quantized,sadhukhan21}.}

\btext{Regarding the symmetry requirements to observe Magnus ballistic transport, we note that the presence of crystalline symmetries such as in-plane $C_2$ and out-of-plane mirror can generate clean MH responses, nullifying other trivial linear transverse signals \cite{Xiao2021}. Using these crystalline symmetries, it is also possible to categorize non-linear anomalous Hall and Magnus Hall effects in different classes of material, in which time-reversal symmetry is preserved but inversion is broken. To be more precise, the noncentrosymmetric point groups containing $\{C_{2z},C_{4z},C_{6z},S_{4z}\}$ symmetry operations force the BC to become zero, and as a consequence both MH conductivity and non-linear Hall conductivity both vanish in these systems. Following the above analysis, crystallographic point groups $\{C_1, C_{1h}, C_{1v}, C_{2}, C_{2v}, C_{3}, C_{3h}, C_{3v}, D_{3h}, D_{3}\}$ (2D transition metal dichalcogenides such as, WTe$_2$, MoS$_2$) and $\{O,T,T_h\}$ (particularly for 3D) allowing non-zero local BC can lead to both MHE and non-linear Hall effect in general \cite{sodemann15,Xiao2021}.}

We would also like to point out that at least one symmetry between TRS and inversion symmetry has to be broken in the system.
Although the previous studies only concentrate on TR invariant and inversion symmetry broken systems, one can in principle get Magnus responses even in the absence of TRS as long as the active momentum modes over the Fermi surface have
finite BC. This further motivates us to study TRS broken topological systems in addition to the TRS invariant topological systems. 
Since the strain, warping, and tilt can modify the Fermi surface significantly as well as the BC distribution accordingly, the Magnus responses can show interesting behavior which we discuss in next section.

\textcolor{black}{It is important to note that along with Magnus Hall current, there exists 
a trivial transverse current (regular Hall current) that arises from the transverse velocity anisotropy of the Fermi surface: $\int_{v_x>0} d\bm k ~v_y(\bm k)~ \delta(\epsilon_{\bm k}-\mu)$. However, according to the symmetry analysis given above, in presence of certain mirror symmetry (specifically, mirror plane perpendicular to the current direction) combined with time reversal symmetry, the trivial transverse current can be shown to vanish leaving a finite MH current\cite{papaj19,Xiao2021}.} 
In addition to the trivial Hall current, linear anomalous Hall current induced by the non-trivial BC can appear simultaneously with MHE \btext{ in time-reversal broken systems}. However, in contrast to MHE, linear anomalous Hall response is not a Fermi surface phenomenon and therefore, does not depend on the derivative of the Fermi function. These two effects can be distinguished by looking at their chemical potential dependencies in experiments. \textcolor{black}{We will discuss these issues elaborately in Sec.~\ref{wsm}}.

\section{Results}
\label{results_MHE}

In this section, we discuss the effect of strain and warping on MH and MN conductivities
in inversion symmetry broken but TRS invariant 2D topological systems, namely for 
ML, BL graphene and surface states of TI. We extend our analysis in 3D topological systems WSMs, breaking either TRS or inversion symmetry, to investigate the effect of tilt and non-linearity on MHE and MNE. At the outset, we note that we will often refer MH and MN conductivities together as MH responses.

\begin{figure*}
\includegraphics[width=\textwidth]{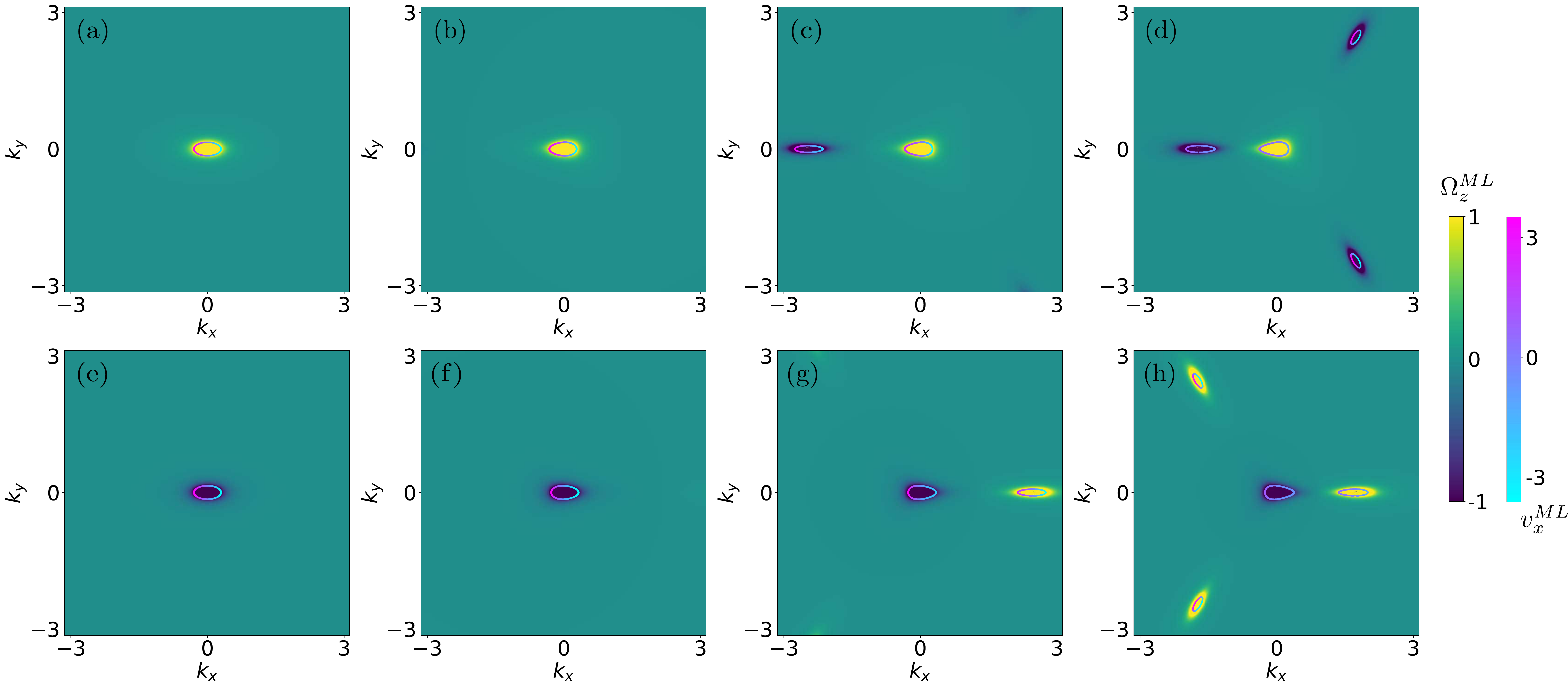}
\caption{ The evolution of BC and Fermi surface in ML graphene
(\ref{ML_tot}) with warping $\lambda_1=\lambda_2=\lambda_3=\lambda$ and strain $v_2 \ne v_1$. Top panel: The BC ($\Omega^{ML}_z(\bm k, \zeta=+1)$) and $x$ component of the velocity ($v_x^{ML}(\bm k,\zeta=+1)$) over the Fermi surface  are shown for (a) unstrained ML graphene without warping ($v_2=v_1$, $\lambda=0$), and strained ML graphene ($v_2=2 v_1$) with $C_3$ symmetric warping of different strengths (b) $\lambda=0.2$, (c) $\lambda=0.35$, (d) $\lambda=0.5$ eV$\cdot$\AA$^2$.
Bottom panel: (e)-(h) depict $\Omega^{ML}_z(\bm k,\zeta=-1)$ and $v_x^{ML}(\bm k,\zeta=-1)$. The strength of $\Omega^{ML}_z(\bm k,\zeta)$ and $v_x^{ML}(\bm k,\zeta)$ are represented in the color codes side by side.  The parameters (in the units of energy eV) used in the calculations are $\Delta_{g}=0.06$eV, $v_1=0.87$eV$\cdot$\AA. The  Fermi surface is plotted for the constant energy $E$=$-0.28$ eV. 
}
\label{fig:BC1}
\end{figure*}


\subsection{Strained monolayer graphene}

The graphene hosts gapless Dirac cones, located at the high symmetry points $\mathbf{K}$ and $\mathbf{K}^\prime$ in the Brillouin zone, with vanishing BC. A finite BC is generated by breaking the inversion symmetry  that can be engineered by placing the graphene sheet on  hBN substrate\cite{Woods2014, Van_2007}. This actually reduces the point group of the system from C$_{6v}$ to C$_{3v}$, and opens up a gap at the Dirac nodes. Upon applying  a uniform uniaxial strain along one of the two main crystallographic directions in graphene, the massive Dirac nodes become shifted from the high symmetry points along the $k_y=0$ line  due to   the combination of TRS and mirror symmetry. The application of the strain creates a difference between hopping amplitudes along the two main crystallographic
directions and therefore, changes the corresponding Fermi velocities\cite{Juan12, Battilomo19}. 

Considering only first-order momentum-strain coupling, the low-energy Hamiltonian of the ML strained graphene can be written as \cite{Battilomo19}

\begin{equation}
\begin{split}
\mathcal{H}_{0}^{ML}({\bm k}) =  \textcolor{black}{\frac{\Delta_{g}}{2}} \sigma_{z} + \zeta v_1 k_{x} \sigma_{x} + v_2 k_y \sigma_{y}, 
\end{split}
\label{ML_Strain}
\end{equation}
where $v_1$ and $v_2 \ne v_1$ are two strain-dependent Fermi velocities along $x$ and $y$ directions respectively, \textcolor{black}{$\Delta_{g}$ is the gap (also called Semenoff mass)}, $\zeta$ is the valley index and $\sigma$'s represent Pauli matrices incorporating sublattice degrees of freedom. To introduce the warping effect in the system, we add a trigonal warping terms, proportional to $k^2$, in the Hamiltonian (\ref{ML_Strain}). Even though the magnitude of the warping term is smaller compared to the leading order term in $k$, it plays a crucial role in MH responses. Now the complete Hamiltonian for the ML graphene in the presence of both uniaxial strain and trigonal warping reads as \cite{Battilomo19};

\begin{equation}
\begin{split}
\mathcal{H}^{ML}({\bm k})  & =  \frac{\Delta_{g}}{2} \sigma_{z} + \zeta v_1 k_{x} \sigma_{x} + v_2 k_y \sigma_{y}+ 2 \zeta \lambda_{3} k_x k_y \sigma_{y} \\
& +(\lambda_{1} k_y^{2} - \lambda_{2} k_{x}^{2}) \sigma_{x}={\mathbf{N}_{\bm k}} \cdot  {\bm \sigma} 
\end{split}
\label{ML_tot}
\end{equation}
with $\mathbf{N}_{\bm k}=\{ \mathbf{N}_{1 \bm k}, \mathbf{N}_{2 \bm k},\mathbf{N}_{3 \bm k} \} =\{\zeta v_1 k_{x}+(\lambda_{1} k_y^{2} - \lambda_{2} k_{x}^{2}),v_2 k_y+2 \zeta \lambda_{3} k_x k_y, \frac{\Delta_{g}}{2}\}$ and $\bm \sigma =\{\sigma_1,\sigma_2,\sigma_3 \}$. Here, $\lambda_1$, $\lambda_2$ and $\lambda_3$ are the warping terms. The energy dispersion of the Hamiltonian for $\zeta=\pm 1$ valley is given by 
\begin{equation}
\begin{split}
E^{ML}({\bm k}) 
&= |\mathbf{N}_{\bm k}| =\pm \sqrt{  \mathbf{N}^2_{1\bm k} + \mathbf{N}^2_{2 \bm k} + \mathbf{N}^2_{3\bm k} }
\end{split}
\label{energy_ML}
\end{equation}
where $+(-)$ represents conduction (valence) band.
The warping results in non-linearity
and strain causes anisotropy in the dispersion.
The BC reads as
\textcolor{black}{
\begin{eqnarray}
\Omega_a({\bm k})=\epsilon_{abc}\frac{\mathbf{N}_{\bm k} \cdot (\frac{\partial \mathbf{N}_{\bm k}}{\partial k_b} \times \frac{\partial \mathbf{N}_{\bm k}}{\partial k_c})}{4 |\mathbf{N}_{\bm k}|^{3}}
\label{Berry_gen}
\end{eqnarray}
}
where $\epsilon_{abc}$ is the usual Levi-Civita symbol.

Since ML graphene is a two-dimensional system, only $z$ component of the BC is nonzero. Using Eq.~(\ref{Berry_gen}), the BC for strained ML graphene ($\zeta=\pm 1$) given in Eq.~(\ref{ML_tot}) can be calculated as

\begin{eqnarray}
&&\Omega^{ML}_{z}({\bm k},\zeta)= \pm \nonumber \\
&& \frac{\Delta_{g}[\zeta v_1 v_2+2k_x(\lambda_3 v_1-\lambda_2 v_2)-4 \zeta \lambda_3(\lambda_2 k_x^2+\lambda_1 k_y^2)]}{4 ( E^{ML}({\bm k}))^3} \nonumber \\
\label{Berry_ML}
\end{eqnarray}
where $+ (-)$ represents conduction (valence) band. Now the $x$ component of the velocity of ML graphene can be written as

\begin{equation}
v^{ML}_{x}({\bm k},\zeta)= \pm \dfrac{2\zeta\lambda_3 k_y \mathbf{N}_{2 \bm k} + \mathbf{N}_{1 \bm k} (\zeta v_1-2\lambda_2 k_x)}{ E^{ML}({\bm k})}.
\label{velo_ML}
\end{equation}

We first consider strained ML graphene $v_1 \ne v_2$ without warping i.e., $\lambda_1=\lambda_2=\lambda_3=0$. It is then clear from the Eq.~(\ref{Berry_ML}) that  the BC at two different valleys ($\zeta=\pm 1$) become opposite of each other for both pristine as well as strained ML graphene i.e., $\Omega^{ML}_{z}(\bm K,\zeta=+1)=-\Omega^{ML}_{z}(\bm K',\zeta=-1)$. Moreover, the BC in this case is directly proportional to the bandgap of the system. On the other hand,  $v_x^{ML}(\bm K,\zeta=+1)=v_x^{ML}(\bm K',\zeta=-1)$ as evident from Eq.~(\ref{velo_ML}).

The above discussion refers to the fact that
there exists equal number of ${\bm k}$-modes for which
$v_x^{ML} >0$ with both
signs of $\Omega(\bm k)$ in each valley. Upon the momentum integration followed by Eqs.~(\ref{sigma})-(\ref{alpha}), the MH and MN conductivities for each valley in ballistic regime can be obtained as 
\ba
 \sigma_{\zeta} &=& \zeta \frac{\Delta_{g} e^2}{8\hbar \pi\mu^2}\Delta U,  \quad  \alpha_{\zeta} = \zeta \frac{\Delta_{g} \pi K^2_B T e}{12\hbar  \mu^3}\Delta U.
 \label{MHE_ml_Strain}
 \ea
Interestingly, the above conductivities are independent of velocities $v_1$ and $ v_2$. This leads to the fact that the uniaxial strain does not affect the MH responses in ML graphene in absence of warping $\lambda_{1,2,3}=0$. Moreover, it is also clear from Eq.~(\ref{MHE_ml_Strain}) that the contributions of MH and MN conductivities for two valleys ($\zeta=\pm 1$) are equal and opposite.  Therefore, summing over the valleys ($\zeta=\pm 1$), the total MH responses vanishes in the strained ML graphene without $\lambda$'s, similar to the case of pristine ML graphene. Instead the valley polarized transport can lead to Magnus valley Hall and  Magnus valley Nernst effects where the electrons with opposite valley index accumulate on the different edges of the sample
\cite{Zeng2012,Mak2012,Xiao07}. To be precise, consider a ML graphene system (with $\lambda_{1,2,3}=0$) with chemical potential ($\mu+\delta \mu/2$), ($\mu-\delta \mu/2$) at the $\zeta=\pm 1$ valleys respectively, the total Magnus valley Hall conductivity will be finite and takes the form $\sigma_{\rm valley}= \sum_{\zeta}\sigma_{\zeta} (\mu + \zeta \delta \mu/2) \simeq
 \frac{\Delta_{g} ~ e^2 \Delta U\delta \mu~}{4\hbar \pi \mu^3}$. Similarly, Magnus valley Nernst conductivity is given by 
$\alpha_{\rm valley}= \sum_{\zeta}\alpha_{\zeta} (\mu + \zeta \delta \mu/2) \simeq 
\frac{\Delta_{g} \pi K^2_B T e \Delta U \delta \mu}{4\hbar  \mu^4}
$. We note that in order to observe the above valley polarized ballistic Magnus response, the dimension of the system has to be smaller or comparable to the mean
free path.

We shall now discuss the effect of C$_3$  symmetric warping in MH responses by considering $\lambda_1=\lambda_2=\lambda_3=\lambda \ne 0$. 
It is clear from the Eq.~(\ref{Berry_ML}) and (\ref{velo_ML}) that  the BC and velocity of each valley are not equal and opposite compared to unwarped case. The evolution of BC and \btext{$v^{ML}_{x}$} for different strength of $\lambda$ are depicted in Fig. \ref{fig:BC1} (b)-(d) and Fig. \ref{fig:BC1} (f)-(h) for $\zeta=1$ and $\zeta=-1$ valleys, respectively. 
The  warping terms introduce satellite Dirac cones with $  \theta \to \theta +  2n\pi/3$ (with $n=1,2,3$) appearing around each Dirac points \cite{Predin16}; here $\theta=\pi$ ($0$) for valleys $\zeta=+1$ ($-1$).  
These additional satellite Dirac cones appear with opposite chirality as compared to that of the parent Dirac cone. The important point to note here is that the relative strength between $v_1$ and $v_2$ determines the distance of satellite Dirac points from the parent Dirac point. For example, $v_1 > v_2$ ($v_1 < v_2$) would cause the satellite Dirac points to move more (less) in $k_x$ direction more (less) than $k_y$ direction.


\begin{figure}[h!]
\includegraphics[width=\columnwidth]{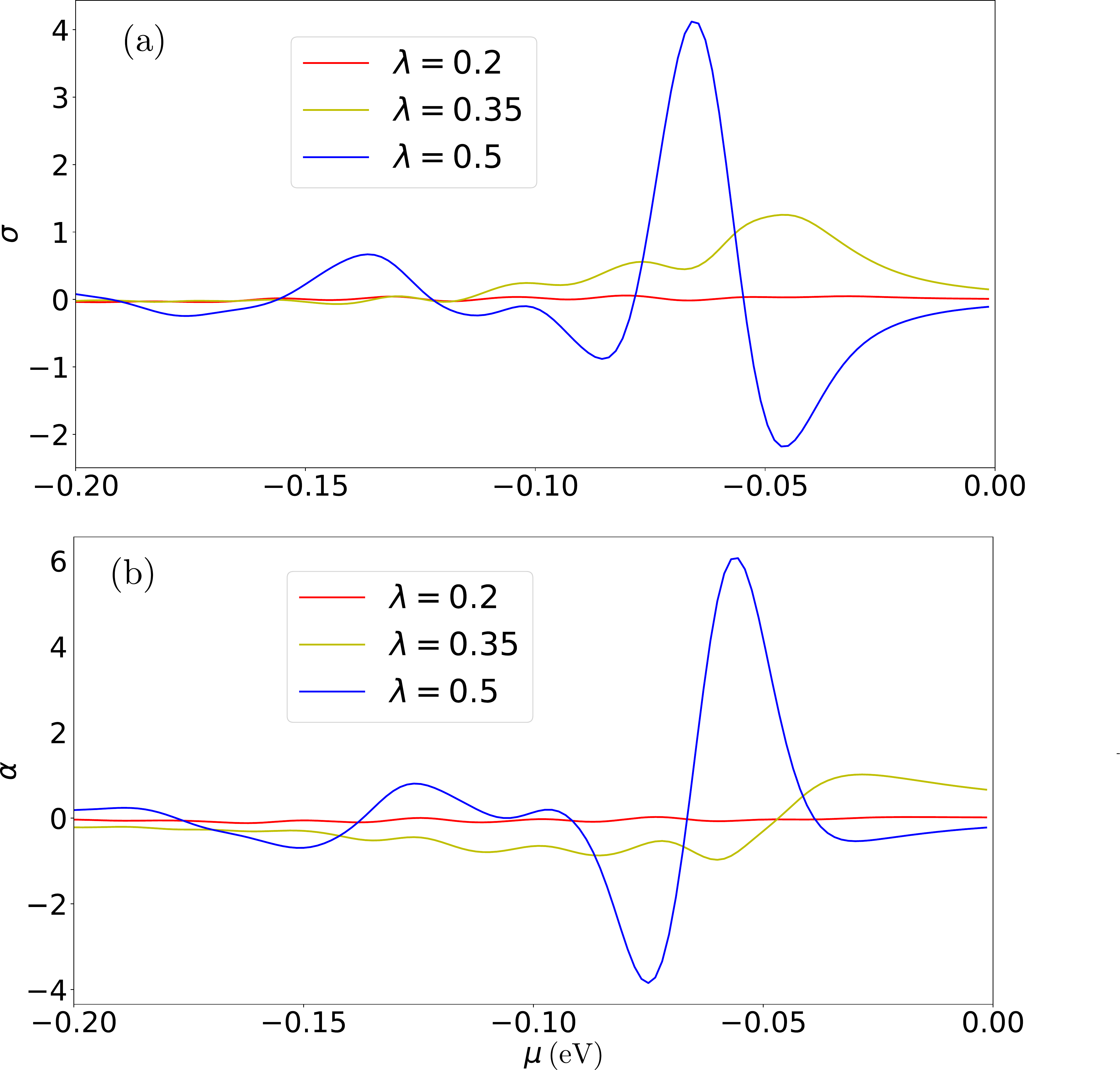}
\caption{The total valley summed contributions of (a) MH conductivity $\sigma$ (in the unit of $10^{-3} e^2/\hbar$) and (b) MN conductivity $\alpha$ (in the unit of $10^{-5} ek_{B}/\hbar$) in the presence of strain ($v_2=2 v_1$) for different warping strengths $\lambda=0.2$, $0.35$ and $0.5$ eV$\cdot$\AA$^2$are shown for ML graphene. Noticeably, the warping can enhance the responses even after valley sum is performed, as it generates asymmetric contributions between valleys. We consider $\Delta U=0.01$ eV and $k_{B}T=0.001$ eV. All other parameters are kept same as that of in Fig.~\ref{fig:BC1}. \textcolor{black}{The chemical potential $\mu$ is chosen in the unit of eV throughout the paper.}}
\label{fig:MHC_ML}
\end{figure}


Following the above discussion, it is evident that each valley of ML strained graphene in the presence of warping does not contribute to MH and MN conductivities in an equal and
opposite manner that we found for strained ML graphene. As a result, the non-zero MH  responses are directly observed by summing over the contribution for both the valleys. The warping  results in non-linear and anisotropic dispersion as shown in Eq.~(\ref{energy_ML}). As a result,  BC becomes anisotropic and exhibits rich features over the Fermi surface (see Fig.~\ref{fig:BC1}). 
Since the expressions of BC and velocity of ML graphene in the presence of warping are quite complicated, it is very difficult to calculate MH and MN conductivities analytically. Therefore, below we have calculated Magnus responses numerically to investigate in detail.

The valley integrated  MH  and MN conductivities as a function of chemical potential ($\mu$) are shown in Fig. \ref{fig:MHC_ML} (a) and (b), respectively.   
We find that the magnitude of both transport coefficients enhances with the increase of warping strength for a fixed strain. The window of  activated momentum modes over the Fermi surface changes with warping strength.  
Moreover, band bending imprints its signature through the Fermi surface distribution. All these lead to the intriguing behavior of the response coefficients. We find that
MH responses show significant different behaviors for $v_1 > v_2$ as compared to $v_2 > v_1$. This refers to fact that the  strain becomes instrumental in controlling the response in presence of  warping terms. We would like to point out that the MH and MN conductivities  can acquire both positive and negative values. This is due to the fact that the  BC around the additional satellite Dirac points take opposite values as compared to that of the for the  parent Dirac point.
 These are the  markedly different responses while the warping terms are introduced in the  ML strained graphene.
Our study further supports that the \btext{Mott relation} can successfully describe  MN conductivity from MH conductivity even in the presence of warping.

\btext{In summary, for monolayer graphene without strain and warping, the MH and MN conductivities for two valleys are equal and opposite, and therefore, summing over the valleys, the total MH responses vanishes \cite{Mandal20}. On the other hand, we show that each valley of monolayer strained graphene in the presence of warping does not contribute to MH and MN conductivities in an equal and opposite manner, and hence lead to a finite valley integrated MH response. This happens because the warping results in non-linear and anisotropic dispersion as shown in Eq.~(\ref{energy_ML}). As a result, BC becomes anisotropic and exhibits rich features over the Fermi surface}.

\subsection{Strained bilayer graphene}.

\begin{figure}
\includegraphics[width=1.5in,height=1.5in]{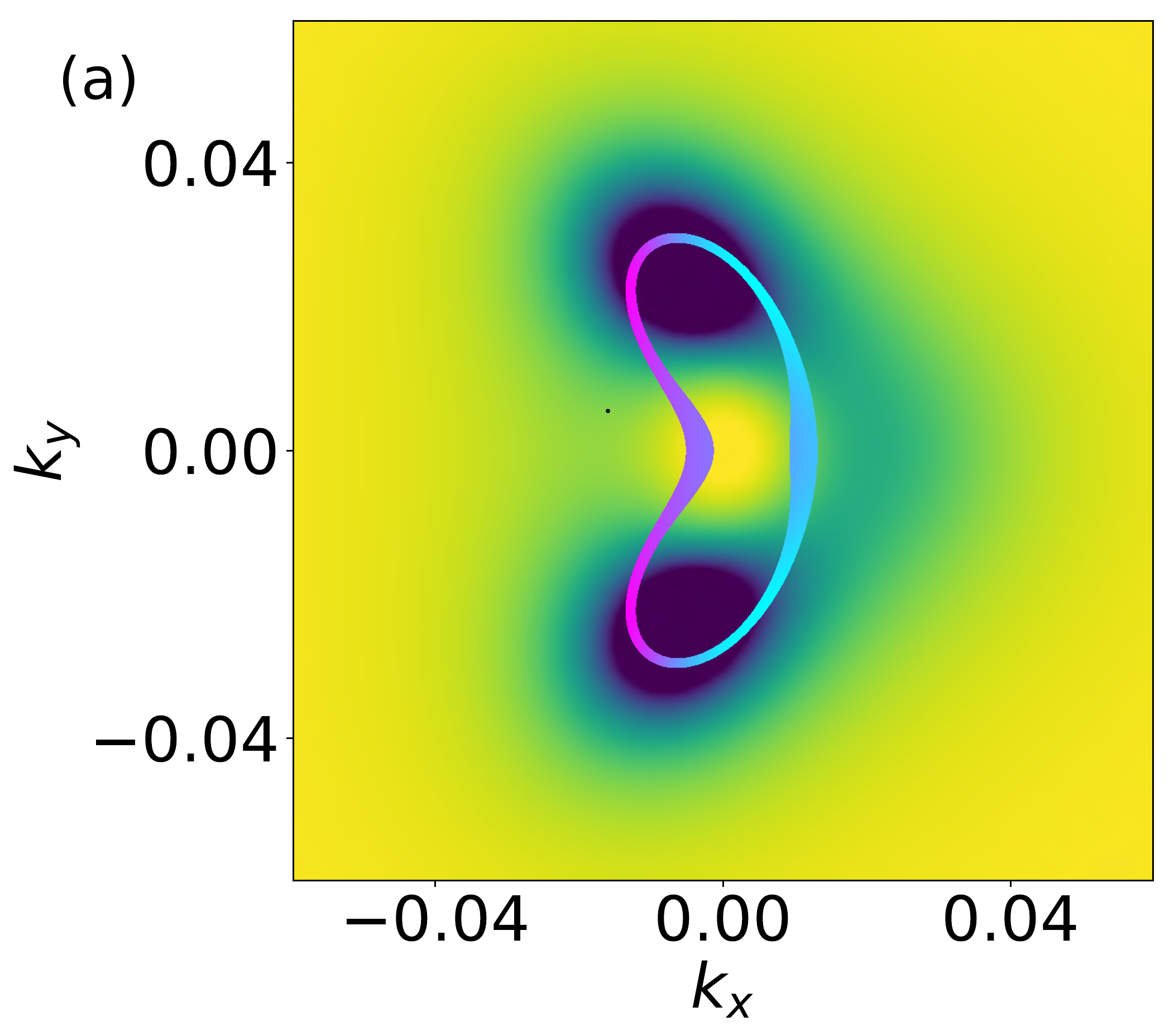}
\includegraphics[width=1.5in,height=1.5in]{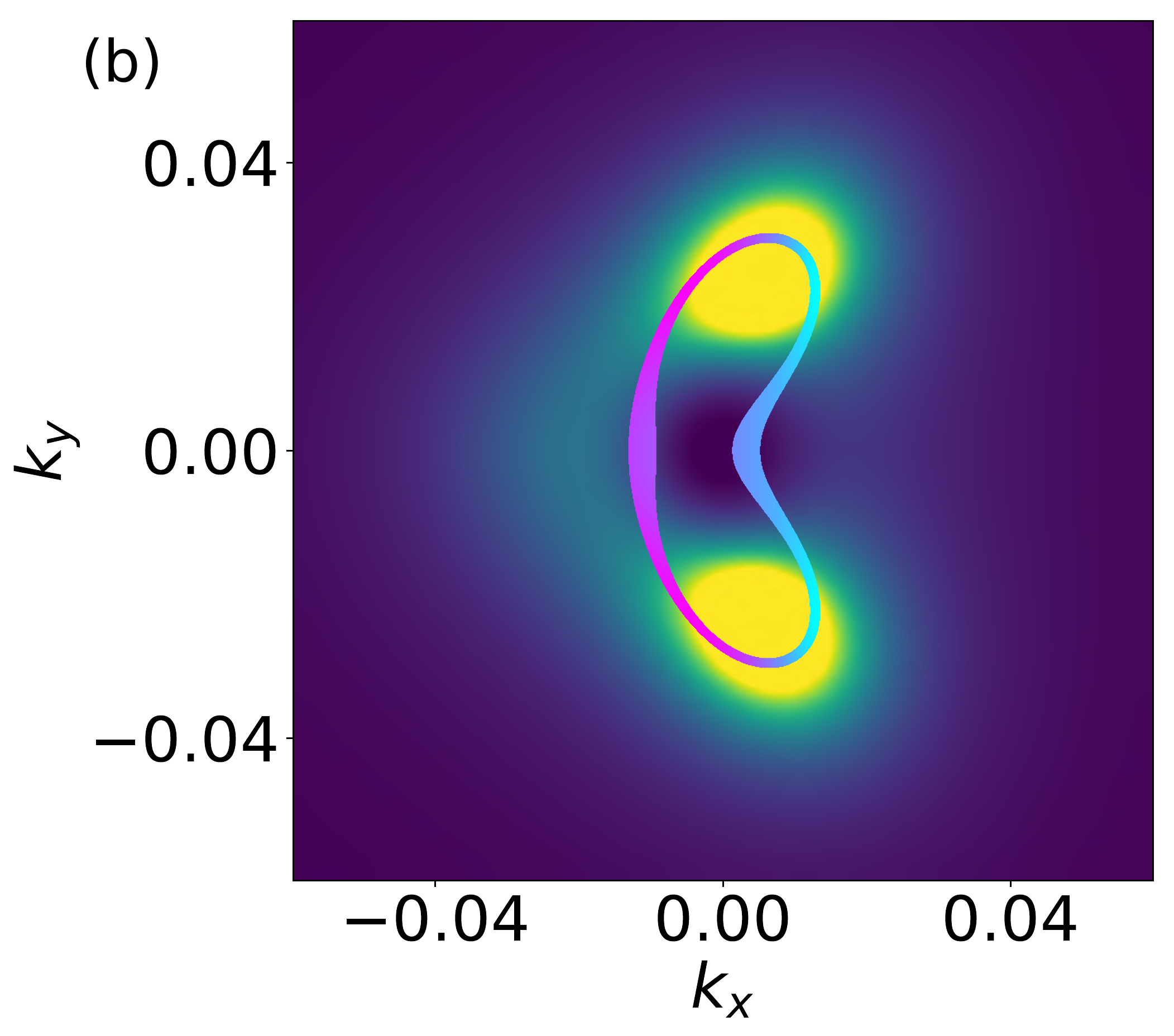}\\
\includegraphics[width=1.5in,height=1.5in]{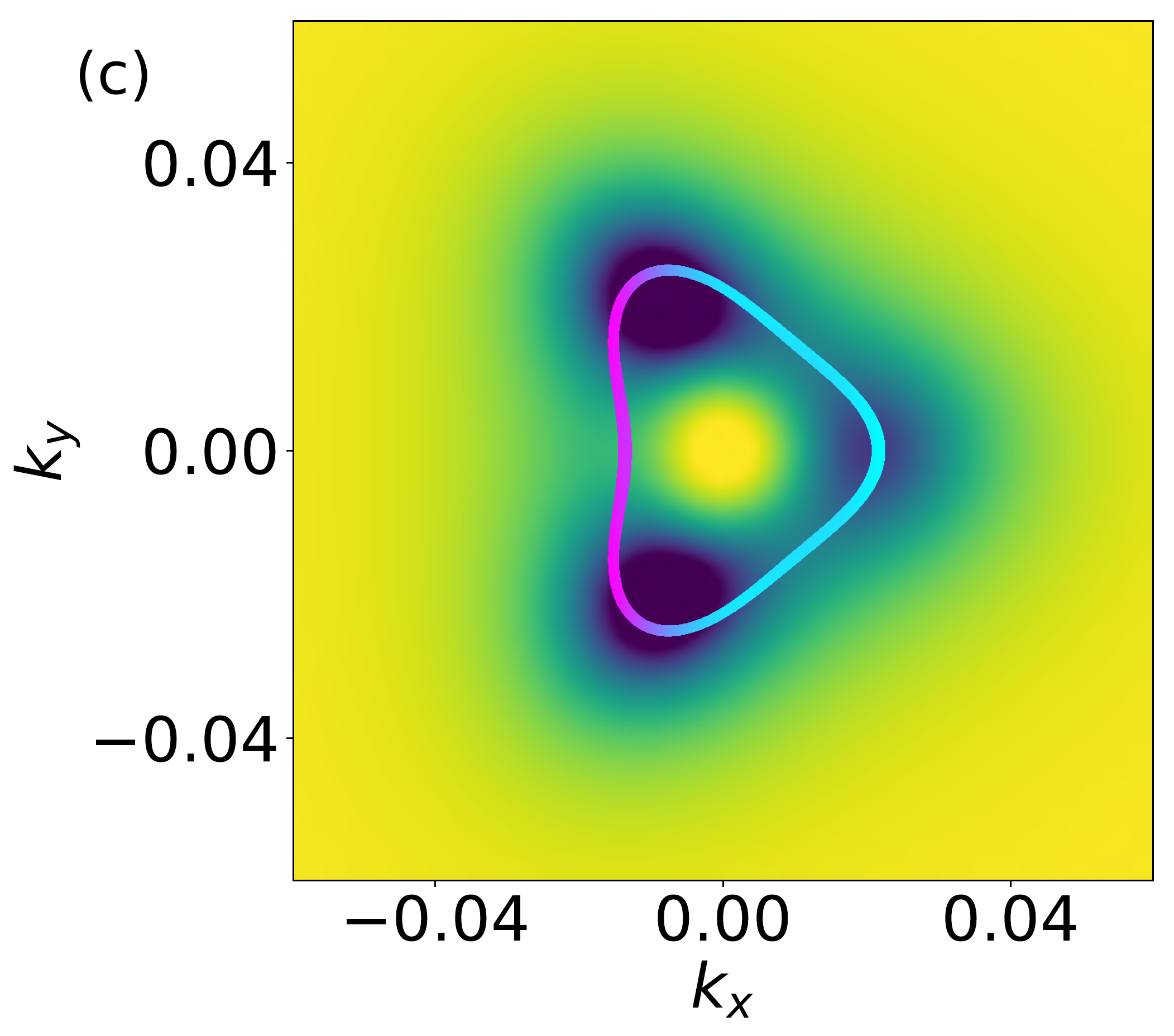}
\includegraphics[width=1.5in,height=1.5in]{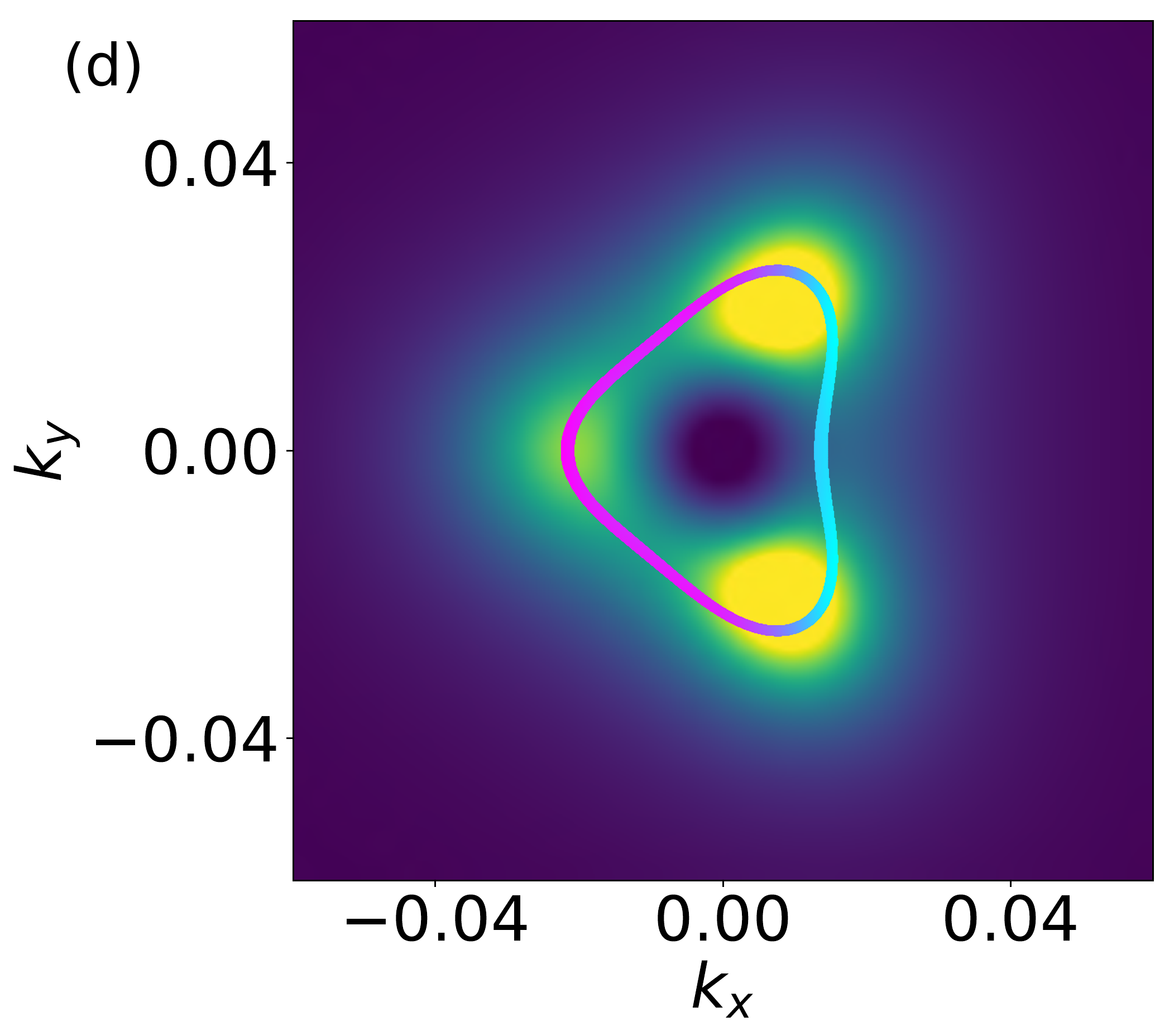}\\
\includegraphics[width=1.5in,height=1.5in]{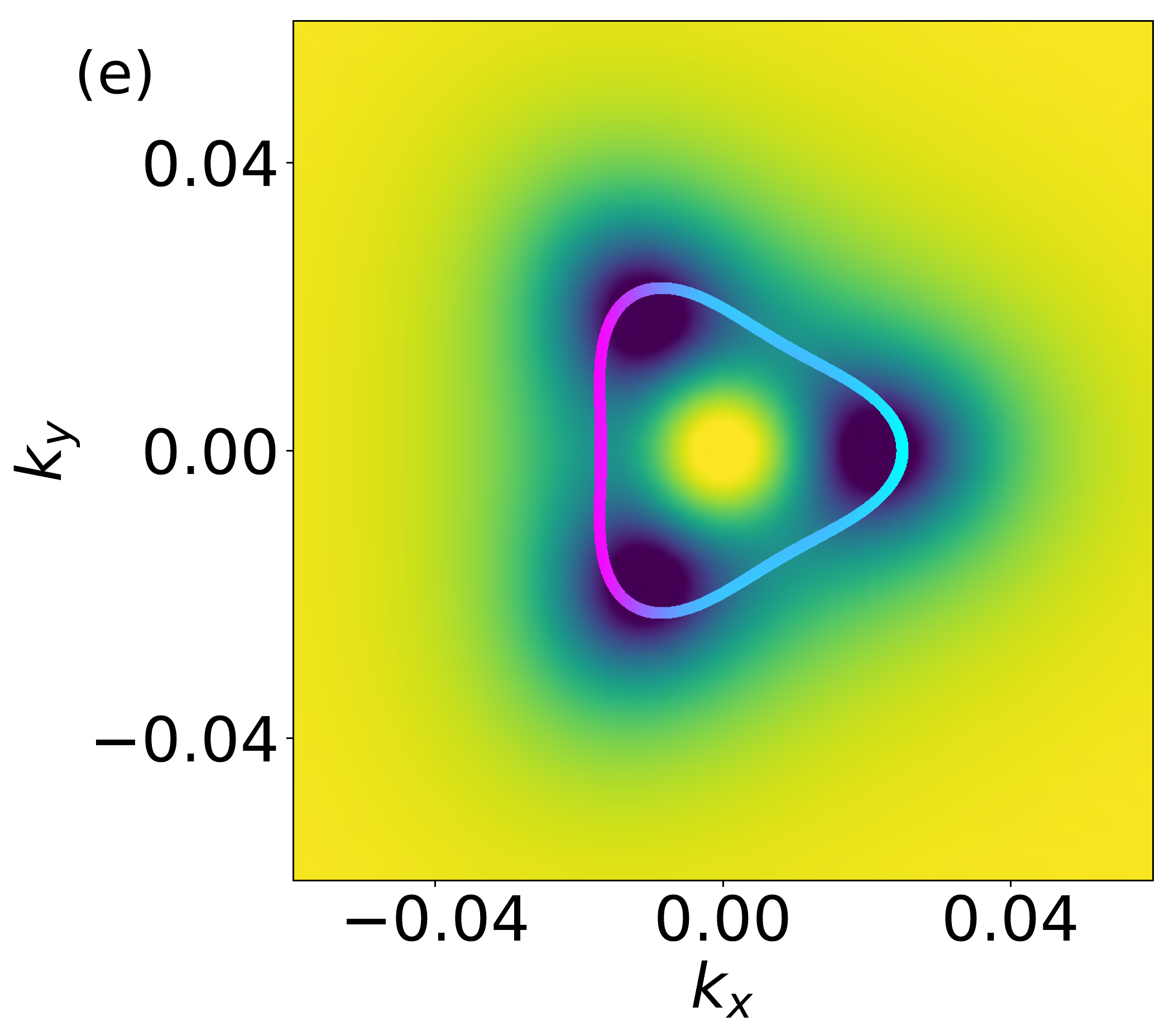}
\includegraphics[width=1.5in,height=1.5in]{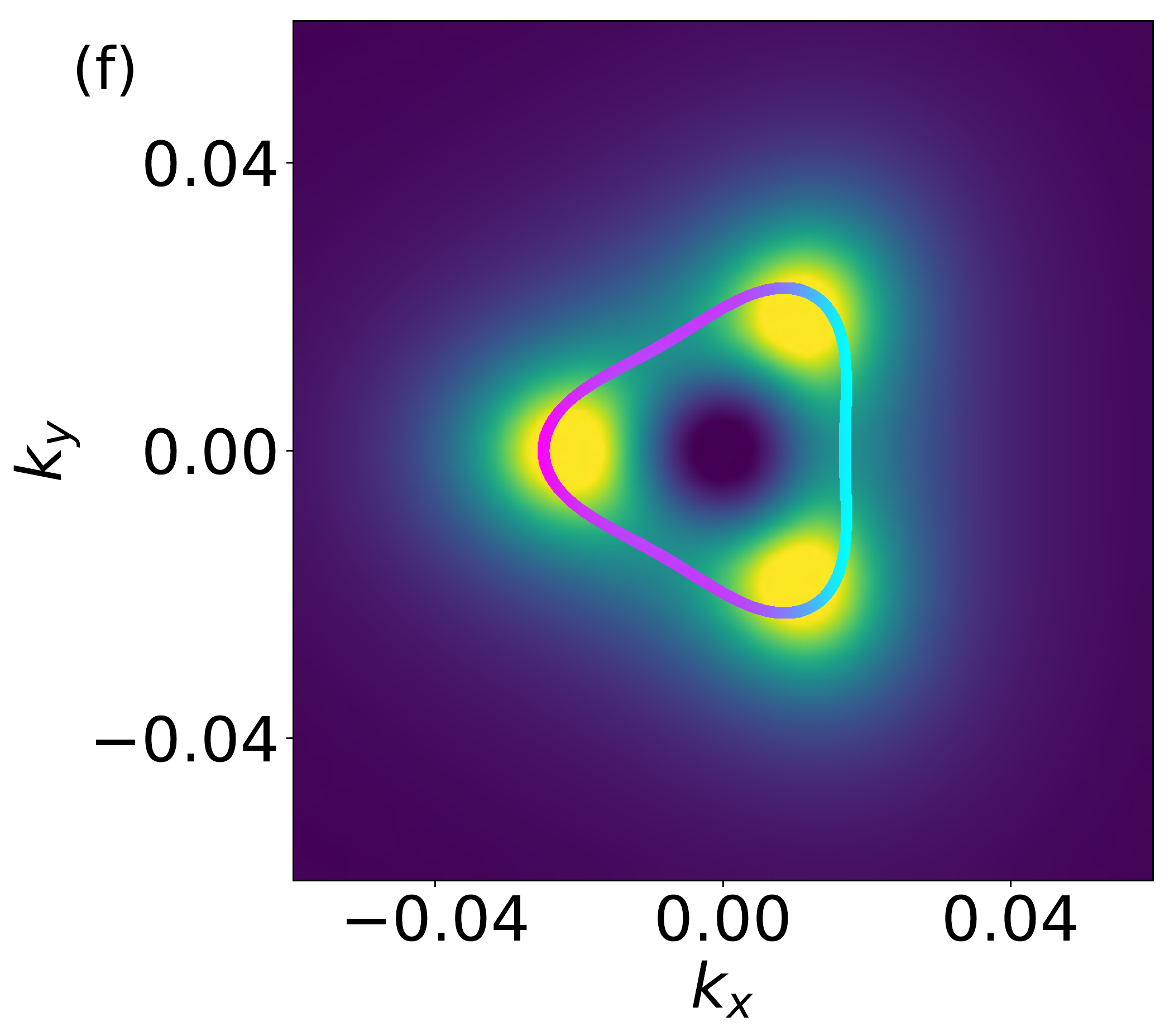}\\
\includegraphics[width=1.5in,height=1.5in]{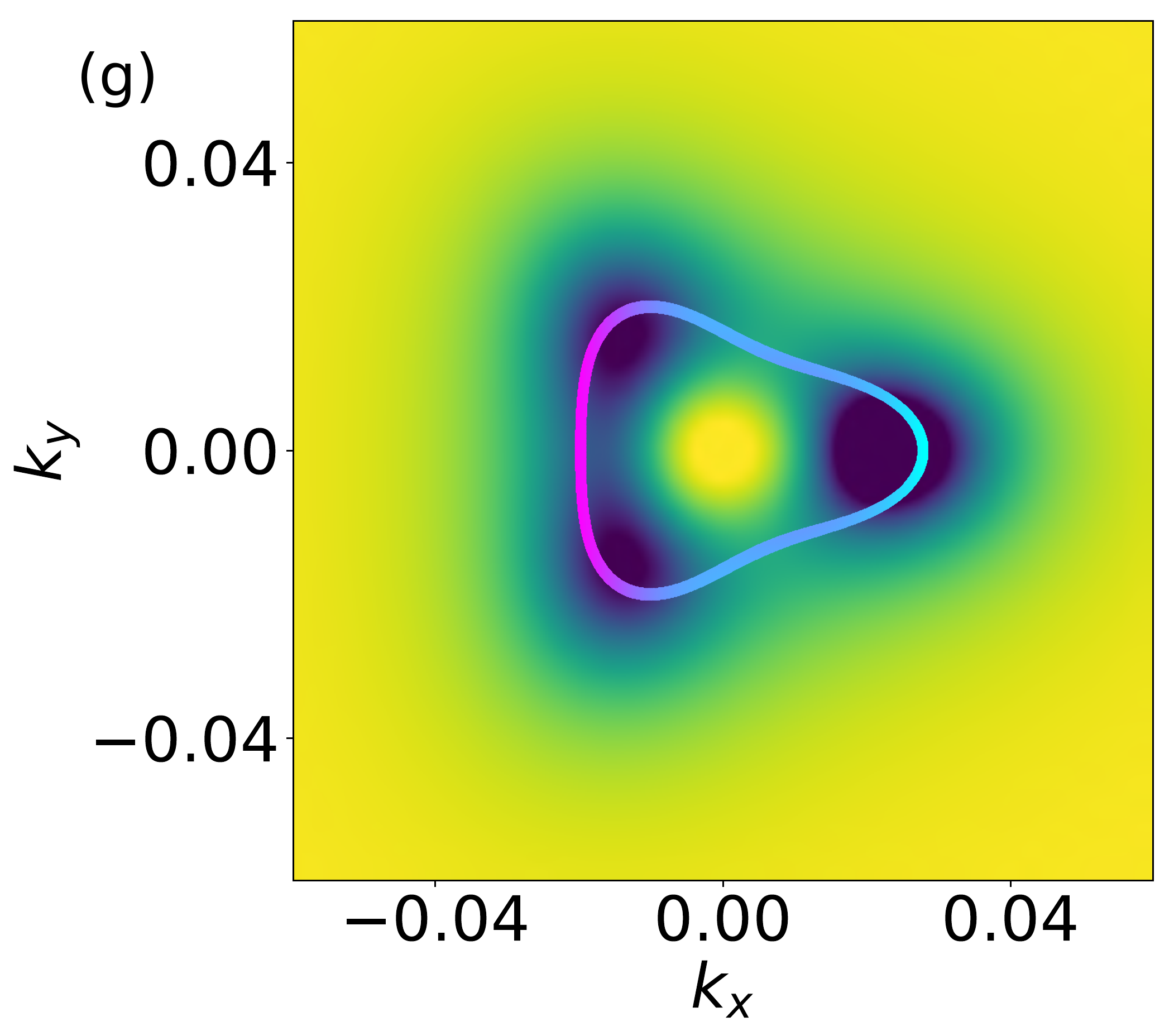}
\includegraphics[width=1.5in,height=1.5in]{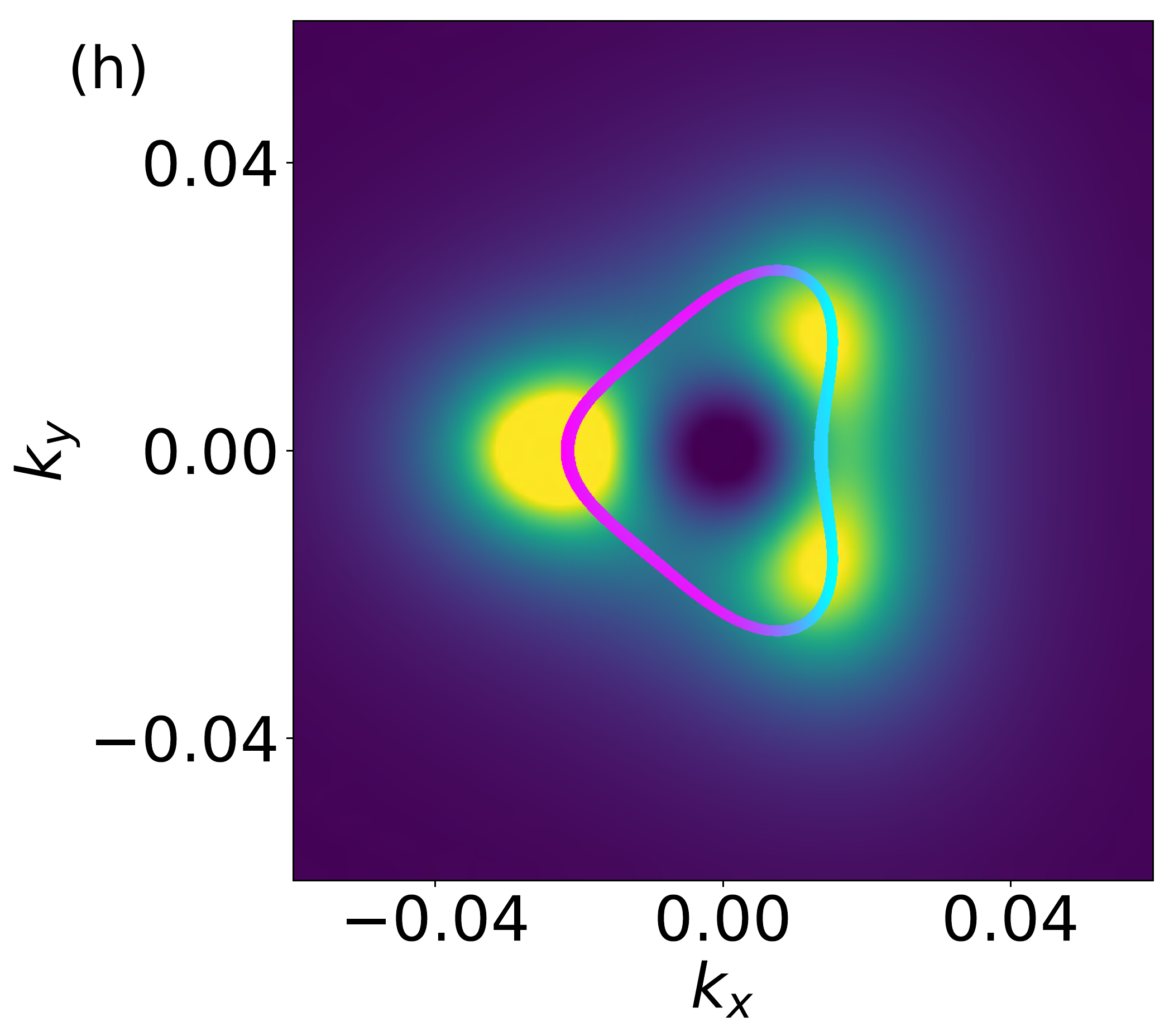}\\
\includegraphics[width=1.5in,height=1.5in]{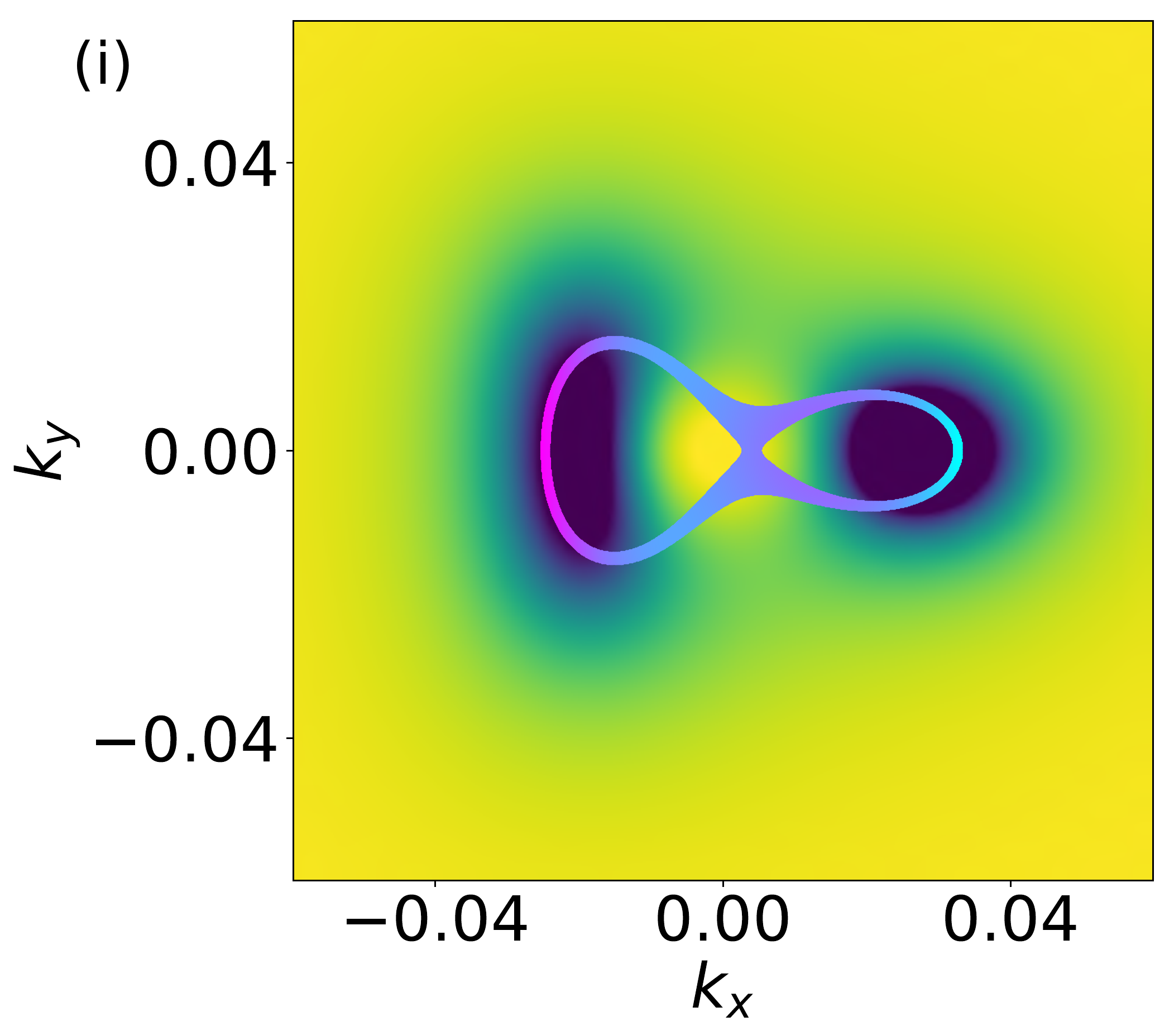}
\includegraphics[width=1.5in,height=1.5in]{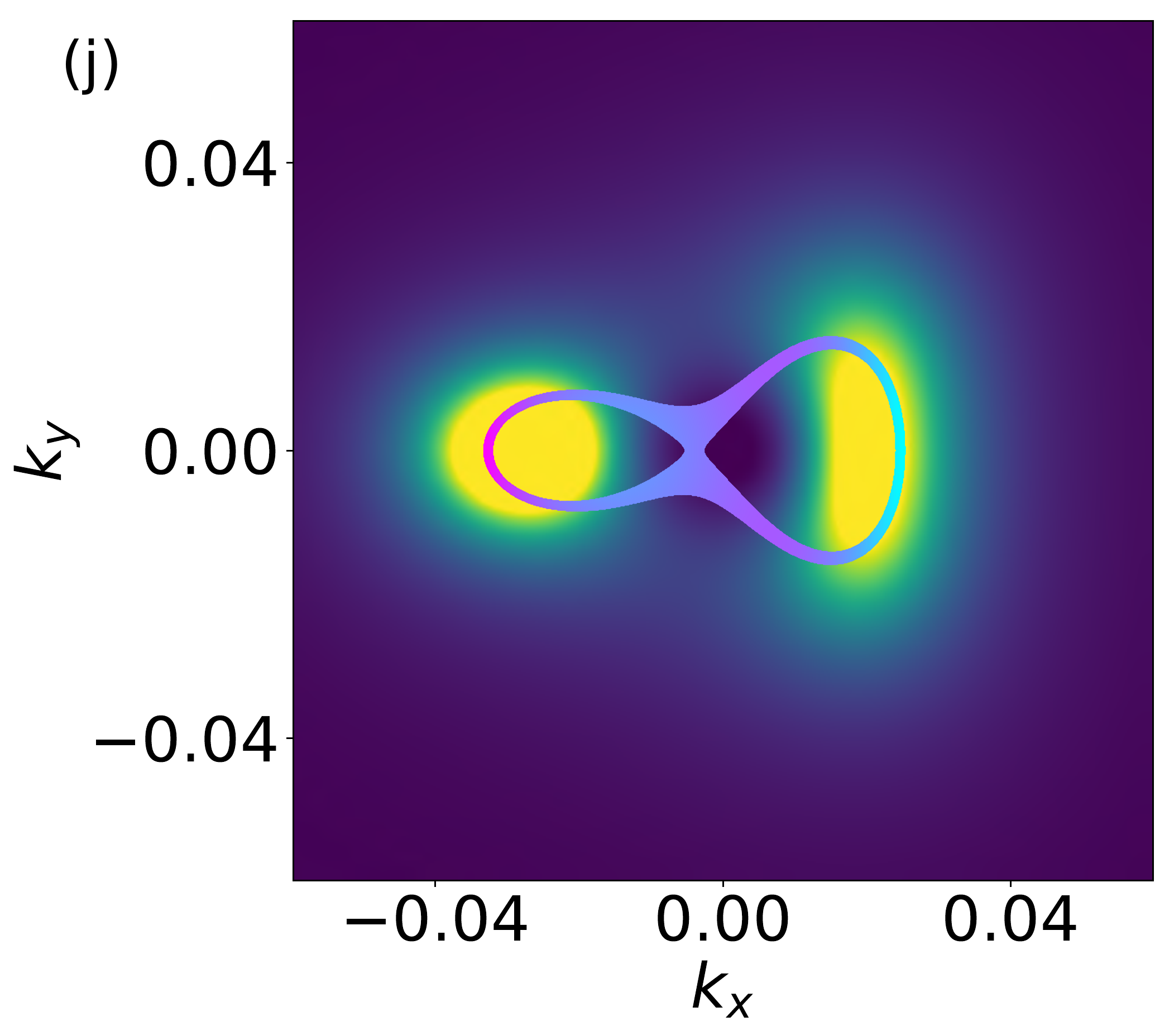}\\
\includegraphics[width=\columnwidth]{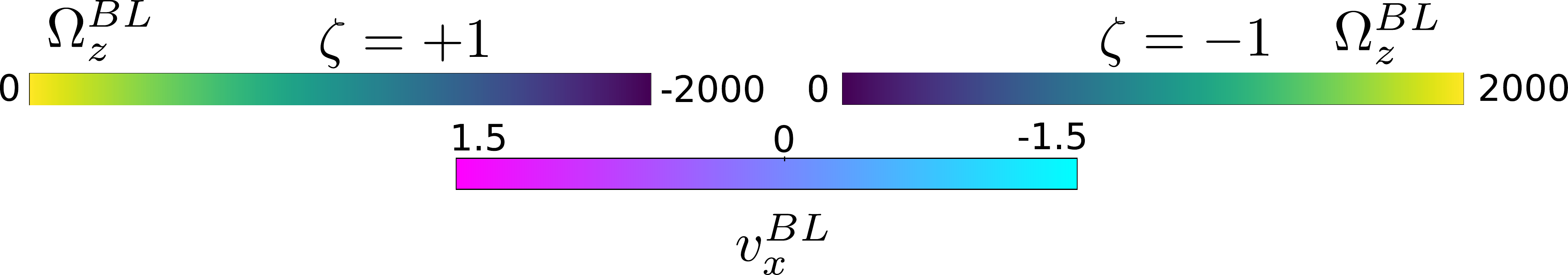}
\caption{  The evolution of BC and Fermi surface in BL graphene (\ref{Ham_BL}) with strain $w$ for warping $\lambda_1=\lambda_2=\lambda_3=\lambda=0.001$eV$\cdot$\AA$^2$.
Left column: The BC $\Omega^{BL}_{z}({\bm k},\zeta=+1)$  and $v_x^{BL}(\bm k,\zeta=+1)$ over the Fermi surface of  BL graphene (a) $w=-3m$, (c) $w=-m$, (e) $w=0$, (g) $w=m$ and (i) $w=3m$  are shown.  
Right column: We repeat the same set of calculations for $\zeta=-1$ valley.
The parameters (in the units of eV) used are $\Delta_{g}=0.06$eV, $m=0.008$eV$\cdot$\AA$^2$, and $v=0.5$eV$\cdot$\AA.  The Fermi surface is plotted for the constant energy  $E$=$-0.04$ eV.
}
\label{fig:BC_BL}
\end{figure}


The BL graphene belongs to the $D_{3d}$ point group symmetry. In order to study the BC mediated transport properties, one has to break the inversion symmetry by applying an external electric field perpendicular to the layers. This reduces the symmetry of the system from $D_{3d}$ to $C_{3v}$, and \textcolor{black}{creates a gap $\Delta_{g}$} as well as finite BC. The application of a uniaxial strain further reduces the symmetry of the point group to $C_{v}$.
The low energy Hamiltonian for an inversion broken BL graphene in the presence of a uniaxial strain is expressed below as \cite{Battilomo19}

\begin{equation}
\begin{split}
\mathcal{H}_{0}^{BL}({\bm k})  & =  \textcolor{black}{\frac{\Delta_{g}}{2}} \sigma_{z}  +\Big(-\frac{1}{2m} (k_y^{2} - k_{x}^{2}) + \zeta v k_{x} + \omega \Big) \sigma_{x}  \\
& - \Big( \frac{1}{m} k_{x} k_{y} + \zeta vk_y \Big) \sigma_{y},
\end{split}
\end{equation}
where $v$ denotes the Fermi velocity related to the skew hopping between the layers, $\zeta$ is the valley index and $m$ represents the effective mass dependent on the coupling between the layers. Here, the effect of strain is coming into the Hamiltonian via $\omega$ ($=A_3-A_0$) where $A_3$ and $A_0$ are pseudogauge fields. Similar to the ML case,  we add  trigonal warping terms to the above Hamiltonian to study its effect. The strained-warped BL graphene Hamiltonian takes the following form 

\begin{equation}
\begin{split}
\mathcal{H}^{BL}({\bm k})  & =  \frac{\Delta_{g}}{2} \sigma_{z}  +\Big(-\frac{1}{2m} (k_y^{2} - k_{x}^{2}) +(\lambda_{1} k_y^{2} - \lambda_{2} k_{x}^{2})\\ 
& + \zeta v k_{x} + \omega \Big) \sigma_{x}
 - \Big( \frac{1}{m} k_{x} k_{y} + \zeta vk_y - 2 \zeta \lambda_{3} k_x k_y \Big) \sigma_{y}\\
 & = {\mathbf{N}_{{\bm k}}} \cdot  {\bm \sigma}
\end{split}
\label{Ham_BL}
\end{equation}
with ${\mathbf{N}_{{\bm k}}} = \{ \mathbf{N}_{1 {\bm k}}, \mathbf{N}_{2{\bm k}},\mathbf{N}_{3{\bm k}} \}=  \{ 
(\lambda_{1} -\frac{1}{2m}) k_y^{2} +
(\frac{1}{2m} - \lambda_{2})
k_{x}^{2} + \zeta v k_{x} + \omega, ( 2 \zeta \lambda_{3} - \frac{1}{m}) k_{x} k_{y} - \zeta vk_y, \frac{\Delta_{g}}{2} \} $. 
The energy dispersion of the Hamiltonian given in Eq.~(\ref{Ham_BL}) can be obtained as
\begin{equation}
\begin{split}
E^{BL}({\bm k}) 
&= |\mathbf{N}_{\bm k}| =\pm \sqrt{  \mathbf{N}^2_{1\bm k} + \mathbf{N}^2_{2 \bm k} + \mathbf{N}^2_{3\bm k} }.
\end{split}
\label{energy_BL}
\end{equation} 

Note that in Eq.~(\ref{Ham_BL}), the warping terms are quadratic in momentum, and hence could be absorbed in the already present quadratic momentum terms which indicate interlayer coupling. In this way, the effective masses associated with $k^2_x$, $k^2_y$ and $k_x k_y$ terms are 
renormalized.  Therefore, we can comment at the outset that  addition of warping might not affect the system substantially as compared to strain in the ML graphene. 


\begin{figure}
\includegraphics[width=\columnwidth]{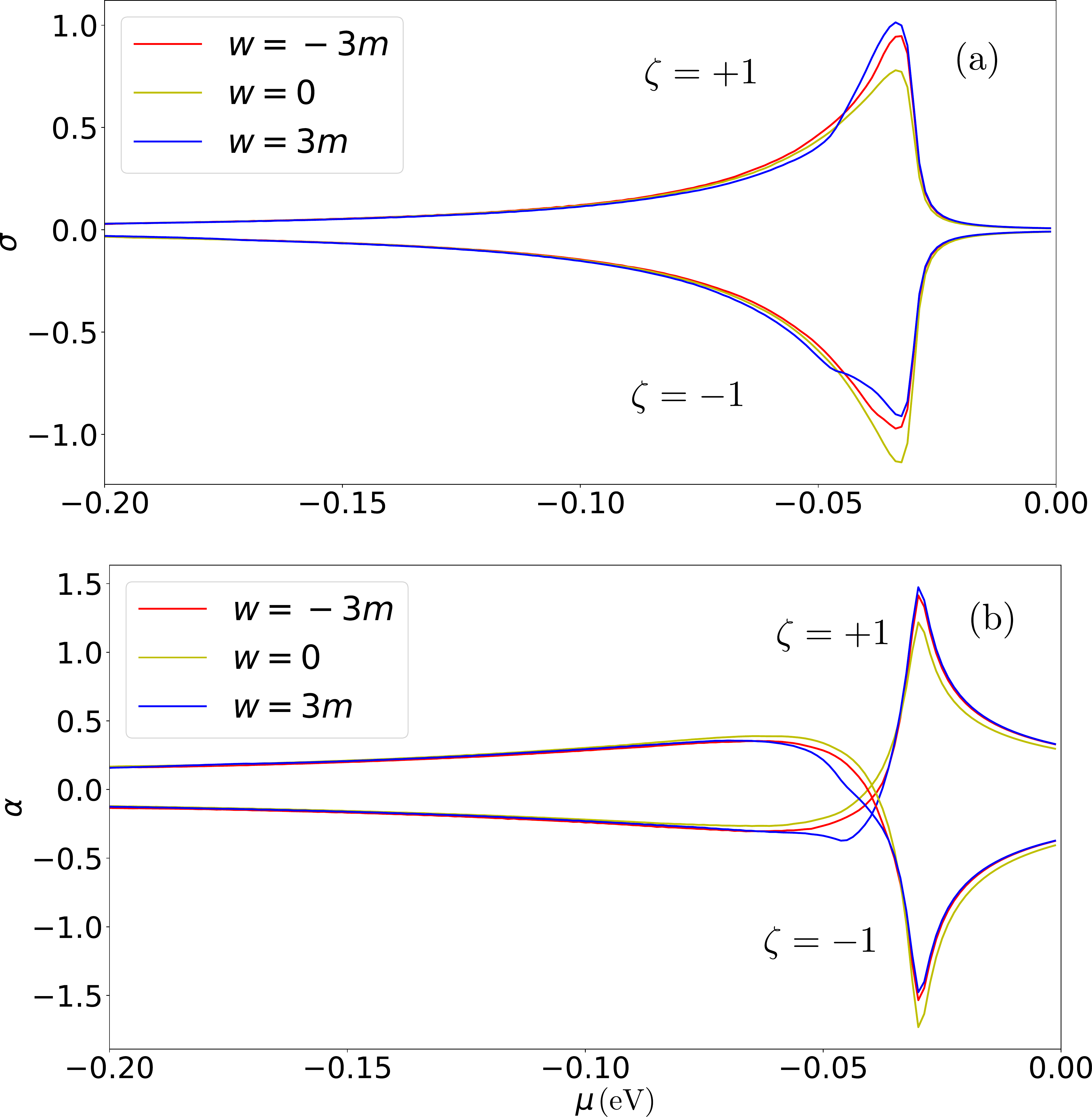}
\caption{Valley responses of (a) MH (in the unit of $10^{2} e^2/\hbar$), (b) MN (in the unit of $10^{-1} ek_{B}/\hbar$)
conductivities in BL graphene with  $\Delta U=0.01$ eV, and $k_{B}T=0.001$ eV for $w=-3m,~ 0,~3m$. All other parameters are kept same as that of in Fig.~\ref{fig:BC_BL}. 
The prominent and asymmetric valley responses in presence of strain for BL graphene are markedly different from the symmetric responses for ML graphene.
}
\label{fig:valley_BL}
\end{figure}


Considering $C_3$ symmetric warping ($\lambda_1=\lambda_2=\lambda_3=\lambda \ne 0$), the BC for strained BL graphene given in Eq.~(\ref{Ham_BL}) can be calculated as
\begin{eqnarray}
&&\Omega^{BL}_{z}({\bm k},\zeta)= \mp \frac{\Delta_{g} \Big[
A(\lambda, k_x) A(\lambda \zeta, k_x) + B(\lambda, k_y)
B(\lambda \zeta, k_y) \Big] }{4(E^{BL}({\bm k}))^{3}} \nonumber \\
\label{Berry_BL}
\end{eqnarray}
with $A(x, k_x)= \frac{k_x}{m}-2 x k_x+\zeta v$, 
$B(x, k_y)= \frac{k_y}{m}-2 x k_y$. 
where $- (+)$ represents conduction (valence) band.
The velocity $v^{BL}_x({\bm k},\zeta)$ takes the form 
\begin{eqnarray}
&&v^{BL}_{x}({\bm k},\zeta)= \pm \frac{ A(\lambda, k_x) 
\mathbf{N}_{1\bm k} - B(\lambda \zeta, k_y)\mathbf{N}_{2\bm k} }{E^{BL}({\bm k})}.
\label{velo_BL}
\end{eqnarray}
A close inspection suggests that the strain factor $w$ does not appear in the numerator of BC for BL graphene unlike the ML graphene where warping factors $\lambda$'s come as corrections over the strain factors $v_1 v_2$. The strain factor comes in the denominator of BC through the energy of the BL graphene. Therefore, the evolution of BC with strain for BL graphene will be significantly different from ML graphene.

The evolution of the $\Omega^{BL}_{z}({\bm k},\zeta)$, $v^{BL}_{x}({\bm k},\zeta)$ and Fermi surface under strain for  valley $\zeta=+1$ and  valley $\zeta=-1$ are shown in left and right column of Fig. \ref{fig:BC_BL}, respectively.
For $w=0$, the three leg gapped Dirac cones (one along $k_y=0$ and the other two symmetrically placed around $k_x=0$ line) are observed at $\theta \to \theta + 2\pi/3$ with $\theta=0$ following the $C_3$ symmetric warping \cite{McCann06}. The BC does not change 
its sign for the three leg gapped Dirac cones within a given valley.
This structure of BC for BL graphene is very different from the ML graphene, where BC around satellite Dirac nodes 
reverses its sign as compared to the parent Dirac node
within a given valley. 
For $w \le -3 m $, there exist two symmetrically placed Dirac cones around $k_y=0$ line, while two cones appears on the $k_y=0$ line for $w \ge 3 m $. The threefold rotational symmetry is
lost in the presence of uniaxial strain. 
Interestingly, the Fermi surface is also deformed for the strained case from its unstrained triangular distribution. To be precise, a singly connected Fermi surface for the unstrained  case
splits into disconnected ones for sufficiently large values of strain. 
The shape and orientation of the Fermi surfaces appear to be different for larger strain in opposite valleys, which leads to  non-identical MH responses in these valleys.


\begin{figure}
\includegraphics[width=\columnwidth]{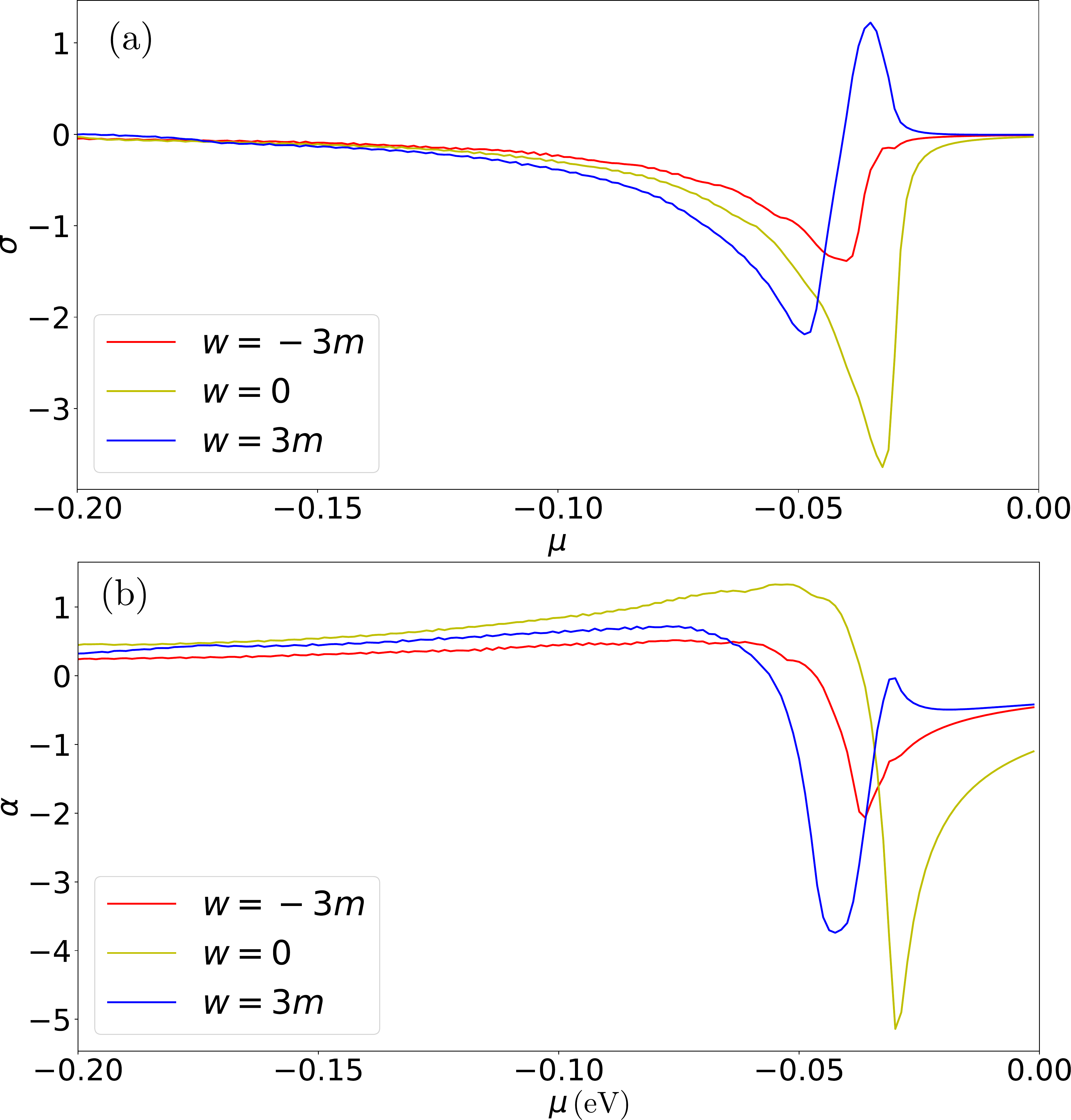}
\caption{Valley summed (a) MH (in the unit of $10 e^2/\hbar$), (b) MN (in the unit of $10^{-2} ek_{B}/\hbar$) conductivities
in BL graphene for $w=-3m,~ 0,~3m$.  The transport behavior changes with strain substantially. We consider the same parameters as used in Fig. \ref{fig:valley_BL}.}
\label{fig:MHC_BL}
\end{figure}


We now investigate the MH and MN conductivities for the individual valleys as shown in Fig. \ref{fig:valley_BL} (a) and (b), respectively. The unstrained case leads to asymmetric response in the valleys which is markedly different from ML strained graphene without warping. Upon inclusion of strain, we find that the positive MH responses 
for the  valley $\zeta=+1$ is more pronounced 
than the negative responses for the 
other valley $\zeta=-1$. This is due to the fact that activated momenta over the Fermi surface do not have exactly opposite BC in terms of their magnitudes and sign.  Moreover, the peak or dip of MHC do not appear at the same chemical potential $\mu$.
The Fermi surface distribution strongly depends on valley as well as strain  explaining the above observation for MH responses. Therefore,  valley polarized transport can in principle be possible like ML strained graphene.

The total valley integrated Magnus responses for
BL graphene are shown in Fig. \ref{fig:MHC_BL} by varying strain parameter $w$. For negative values of strain parameter i.e., $w<0$, MH conductivity always acquires negative values and a dip appears at a certain $\mu$ value. The height of the dip increases and its position moves toward $\mu \to 0$ with decreasing negative strain. On the other hand, by changing the sign of strain i.e., $w>0$, the dip structure of MH conductivity gets bifurcated
into a dip and peak structure. The valley polarized structure of MH responses can explain these observations. We notice qualitatively similar response in MN conductivity.

\btext{In summary, for bilayer graphene, we find that strain enhances asymmetry between the valley polarized contribution, resulting in distinct transport signatures for positive and negative strain as compared to the unstrained bilayer graphene}.

\subsection{Hexagonal warped topological insulator}
\label{hexagonal}


\begin{figure*}[tb]
\includegraphics[width=2\columnwidth]{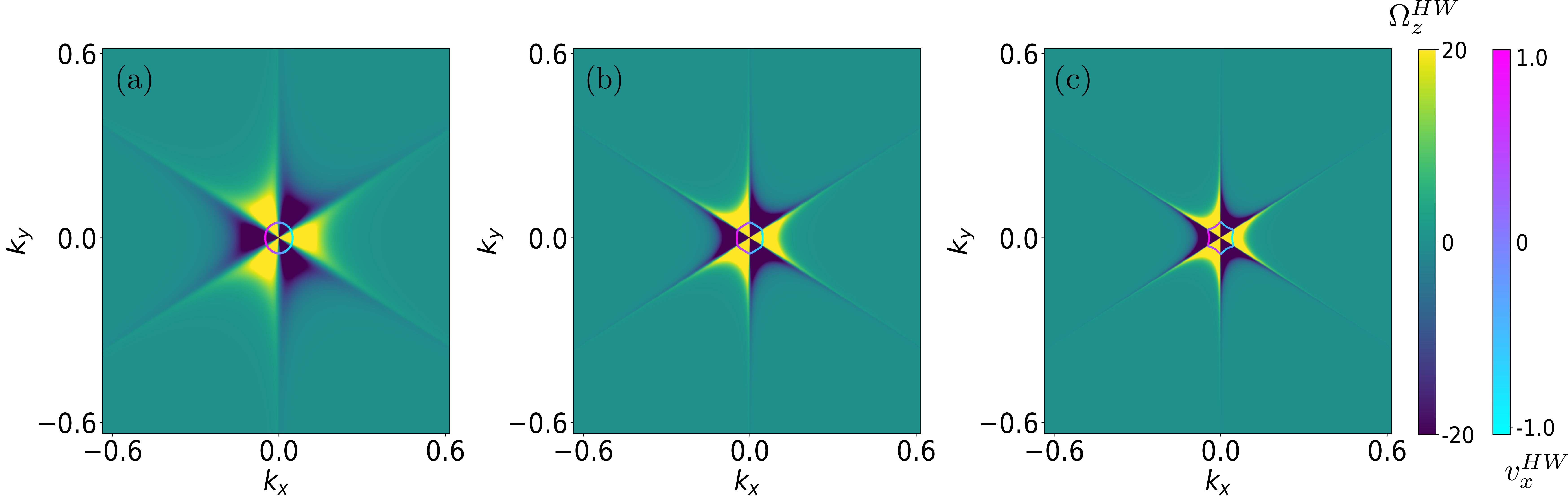}
\caption{Evolution of BC and Fermi surface, 
calculated from Eq.~(\ref{ham_hexagonal}),
with different warping strengths (a) $\lambda$=$50$ eV$\cdot$\AA$^3$, (b) $\lambda$=$200$ eV$\cdot$\AA$^3$ and (c) $\lambda$=$400$ eV$\cdot$\AA$^3$ are shown. Fermi surface is plotted for $E$=$-0.05$ eV. We note that for $\lambda$=$50$ eV$\cdot$\AA$^3$, the Fermi surface remains circular, which gradually evolves to hexagonal shape  with increasing $\lambda$. We consider $v$=$1$ eV$\cdot$\AA  and $E_0$=$0$ in our calculation.}
\label{fig:BC_HW}
\end{figure*}


We consider the two-dimensional surface Hamiltonian of a TRS invariant TI namely, $Bi_2Te_3$, hosting  a unique Fermi surface that encloses an odd number of Dirac cones in the surface Brillouin zone. The spin-orbit coupling that is the linear order term in $k$ leads to the band inversion in this system. The minimal two band model contains cubic terms in $k$ in addition to the linear terms in $k$.
This warping can be considered as a counterpart of cubic Dresselhaus spin-orbit coupling term. We note that hexagonal warping incorporates  one order higher momentum coupling than the  trigonal warping terms. This further allows us to investigate the non-trivial effects of this term  that are not captured by the trigonal warping terms.

Considering the threefold rotation $C_3$ around the $z$ axis and mirror symmetry $M$: $x\to -x$, the low-energy model around the gapless $\Gamma$ point thus  reads \cite{Fu09}
\begin{equation}
 H^{HW}({\bm k})= E_0 ({\bm k}) + v(k_x \sigma_y -k_y \sigma_x) + \frac{\lambda}{2}(k_+^3 + k_-^3)\sigma_z,
 \label{ham_hexagonal}
\end{equation}
with $E_0(k)= \frac{k^2}{2m^*}$ causes the particle-hole asymmetry. Dirac velocity $v$ can be considered $k$ independent without loss of generality. Here $k_{\pm}=k_x \pm i k_y$ and $\lambda$ is the strength of hexagonal warping.

The energy spectrum becomes
\begin{equation}
 E^{HW}({\bm k})=E_0({\bm k}) \pm \sqrt{v^2k^2 +  \lambda^2 k^6 \cos\phi},
\end{equation}
where $\phi={\rm arctan}\frac{k_y}{k_x}$. Using Eq.~(\ref{Berry_gen}), the BC and $x$-component of the velocity for the above Hamiltonian can be obtained as

\begin{eqnarray}
&&\Omega_{z}^{HW}({\bm k})= \pm \frac{\lambda v^2(3k_x k_y^2-2k_x^3)}{2 (v^2k^2 +  \lambda^2 k^6 \cos\phi)^\frac{3}{2}} 
\label{Berry_HW}
\end{eqnarray}
\begin{equation}
v^{HW}_{x}({\bm k})= \frac{k_x}{m^{*}}\pm \dfrac{2v^2 k_x+\lambda^2 k^3 k_y^2 +6 \lambda^2 k^4 k_x \cos \phi}{ 2\sqrt{v^2k^2 +  \lambda^2 k^6 \cos\phi}}.
\label{velo_HW}
\end{equation}
where $+ (-)$ represents conduction (valence) band. The band structure is sixfold symmetric under $\phi \to \phi + 2 \pi/6$. It is clear from the Eq.~(\ref{Berry_HW}) that the BC is zero in the absence of warping. The band structure is sixfold symmetric under $\phi \to \phi + 2 \pi/6$. The BC distribution with different strength of warping parameter is depicted in Fig. \ref{fig:BC_HW}.  The BC always shows a snowflake like distribution irrespective of the strength of warping. However, the BC acquires substantially large value 
around the $\Gamma$ point with increasing $\lambda$ as also suggested from Eq.~(\ref{Berry_HW}). Moreover, it reverses sign between two subsequent interval $ 2n \pi/6 \to 2(n+1) \pi/6 $. On the other hand, the shape of the Fermi surface changes with warping. Specifically, for small warping the Fermi surface takes circular shape. With increasing warping strength, it becomes non-circular with relatively sharp tips extending along high symmetry direction and curves inward in between. Such transformation of Fermi surface  from circular to snowflake would initiate interesting transport behavior which we discuss below.


\begin{figure}
\includegraphics[width=0.95\columnwidth]{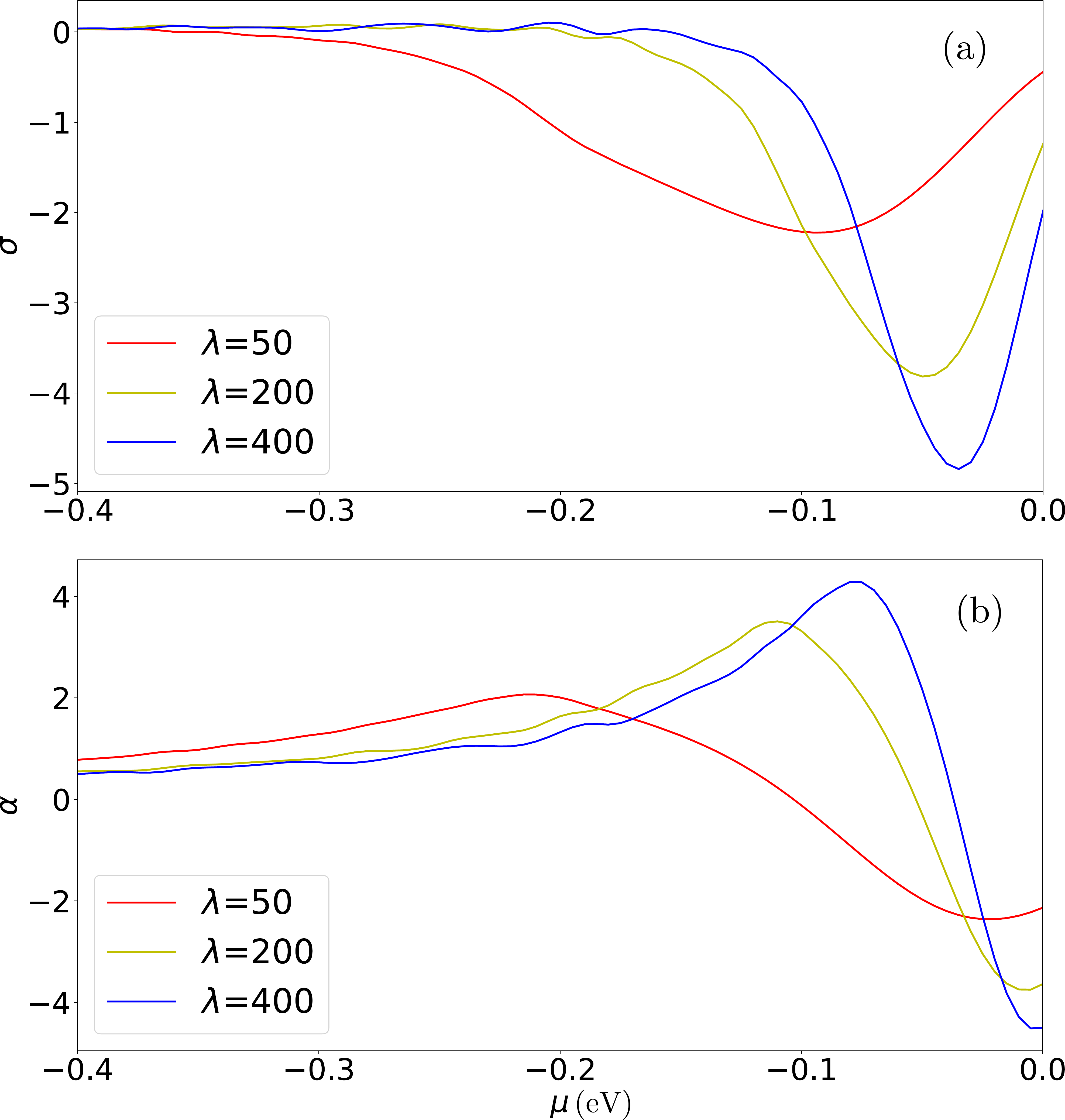}
\caption{(a) MH (in the unit of $10^{-2} e^2/\hbar$) and (b) MN (in the unit of $10^{-4} ek_{B}/\hbar$) conductivities as a function of chemical potential for different warping strength $\lambda=50$, $200$ and $400$ eV$\cdot$\AA$^3$ are depicted. The  parameters used are $v$=$1$ eV$\cdot$\AA, $\Delta U=0.01$ eV, and $k_{B}T=0.001$ eV.}
\label{fig:MHC_HW}
\end{figure}


The MH and MN conductivities as a function of chemical potential for different strengths of warping parameter are shown in Fig. \ref{fig:MHC_HW} (a) and (b), respectively. We find that the magnitude of the  MH responses are increasing as well as become more pronounced and sharp with the increase of warping strength. In addition, the position of the dips (peaks) in MH (MN)  conductivities moves toward $\mu=0$ with increasing $\lambda$. 
Moreover, the negative sign of MH conductivity appears because of the majority of negative BC over the activated momentum modes in the Fermi surface.

\subsection{Weyl semimetals}
\label{wsm}


\begin{figure*}[tb]
\includegraphics[width=2.0\columnwidth]{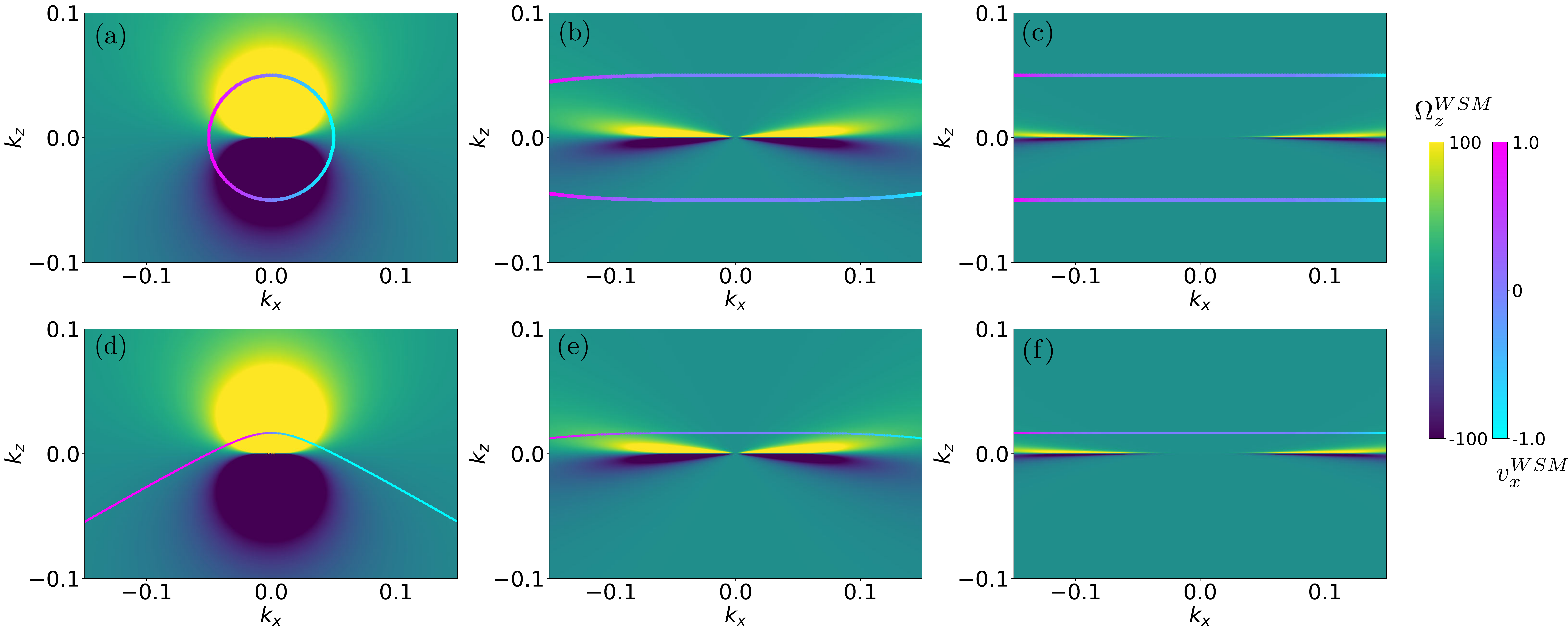}
\caption{The distribution of the BC and the Fermi surfaces, 
calculated from Eq.~(\ref{h1}),
in (a) [(d)] single Weyl node with $n=1$, (b) [(e)] double Weyl node with $n=2$ and (c) [(f)] triple Weyl node with $n=3$ for untilted case i.e., $C_{+}=0$ [tilted case i.e., $C_{+}=2.0$]  are shown. The Fermi surface is calculated for $E$=$-0.05$ eV with $v=1$ eV$\cdot$\AA. The deformation of BC is clearly observed with increasing non-linearity and anisotropy in the WSM.}
\label{fig:BC1_WSM}
\end{figure*}

Going beyond the 2D systems, we will now calculate MH responses in 3D WSMs which can be thought of as a 3D analogue of graphene \cite{Yan17}. The low-energy effective Hamiltonian describing the Weyl node with topological charge $n$ and chirality $\zeta$ is written as~\cite{Gxu11, bernevig12,Yang2014, Roy_2017} 
\begin{equation}
 \label{h1}
H_{\bm k}^{\zeta}=  C_{\zeta}( k_z-\zeta Q) + \zeta  \alpha_n {\boldsymbol{\sigma}} \cdot ({\mathbf n}_{\bm k}- \zeta {\mathbf e}).
\end{equation}
where $\zeta=\pm 1$, $k_{\bot}=\sqrt{k_x^2+k_y^2}$, $\phi_k={\rm arctan}(k_y/k_x)$ and ${\mathbf e}=(0,0,Q)$. The Weyl nodes of opposite chirality are shifted by an amount $\pm Q$ in momentum space due to broken TRS. $C_{\zeta}$ indicates the tilt parameter associated with
Weyl node  with chirality $\zeta$. Here, $\bm \sigma= \{ \sigma_x,\sigma_y,\sigma_z \}$  and $\mathbf{n}_{\bm k}= \{ \alpha_{n} k^n_{\bot}\cos \left( n \phi_{k} \right), \alpha_{n} k^n_{\bot}\sin \left( n \phi_{k} \right), v k_z \}$. 
The factor $\alpha_n$ bears the connection to the Fermi velocity.  $v$ is equivalent to the velocity associated with $z$-direction.
For the sake of simplicity, we consider $Q=0$ and take into account the Weyl nodes of opposite chirality separately. For $C_{\zeta}=0$,
electron and hole bands touch at the Weyl point leading to a
point-like Fermi surface. When the magnitude of the tilt parameter is
small enough i.e., $|C_{\zeta}|/v\ll 1$, the Fermi surface is still point-like, and is characterized as the type-I Weyl node. With the increase of $C_{\zeta}$, electron and hole pockets now
appear at the Fermi surface for $|C_{\zeta}|/v \gg 1$ leading to a distinct phase, which
is designated as a type-II Weyl node. In this work, we consider two types of tilt configuration for the WSM: i) chiral tilt  i.e., $C_+=-C_-$, and ii) achiral tilt  i.e., $C_+=C_-$.

Now the energy dispersion of the multi-Weyl node with $\zeta=+1$ is given by 
\begin{equation}
E^{WSM}({\bm k},\zeta)= C_{\zeta} k_z \pm \epsilon_{\bm k} 
\label{eq_multi2}
\end{equation}
where $
\epsilon_{\bm k} =  \sqrt{ \alpha^2_{n} k^{2 n}_{\bot} + v^2 k^2_z} $
and  $+(-)$ represents conduction (valence) band. It is now clear that the dispersion around a Weyl node with $n=1$ is isotropic in all momentum directions. On the other hand, for $n>1$, we find that the dispersion 
around a double (triple) Weyl node becomes quadratic (cubic) along both $k_x$ and $k_y$ directions whereas varies linearly with $k_z$.

Using Eq.~(\ref{Berry_gen}), the explicit form of $z$-component BC  associated with the multi-Weyl node can be written as
\begin{equation}
{\Omega}^{WSM}_{z}({\bm k},\zeta) =\pm \frac{1}{2} \frac{\zeta n^2 v \alpha_n^2 k^{2n-2}_{\bot} }{\epsilon_{\bm k}^{3}}
k_z.
\label{eq_bcl}
\end{equation}
It is clear from the Eq.~(\ref{eq_bcl}) that, similar to energy dispersion, the BC is isotropic in all momentum directions for single WSM  whereas it becomes anisotropic for WSMs with $n>1$ i.e., for double WSM ($n=2$) and triple WSM ($n=3$) due to the presence of $k^{2n-2}_{\bot}$ factor and topological charge $n$. Moreover, the BC reverses its  sign, retaining the  magnitude same,
for Weyl nodes of opposite chiralities $\Omega^{WSM}_{z}(\bm k,\zeta=+1)=-\Omega^{WSM}_{z}(\bm k,\zeta=-1)$. The behavior of BC for both untilted ($C_\zeta=0$)  and  tilted 
($C_\zeta \ne 0$)
multi-Weyl node with $\zeta=+1$  are shown in Fig. \ref{fig:BC1_WSM}. With increasing the topological charge $n$, the single positive and negative lobe of BC gets divided into one pair of lobes. These lobes for $n=2$ and $3$ are deformed with respect to that of $n=1$. The separation between these lobes increases over the $k_z=0$ line.  The BC changes sign with the chirality of the Weyl node for all WSMs as indicated in  Eq.~(\ref{eq_bcl}). However, the BC is identical in both tilted and untilted WSMs because the tilt does not effect BC. On the other hand, in contrast to BC, the Fermi surface drastically changes in tilted WSM compared to untilted WSM as shown in Fig. \ref{fig:BC1_WSM}.


\begin{figure}[h!]
\includegraphics[width=0.95\columnwidth]{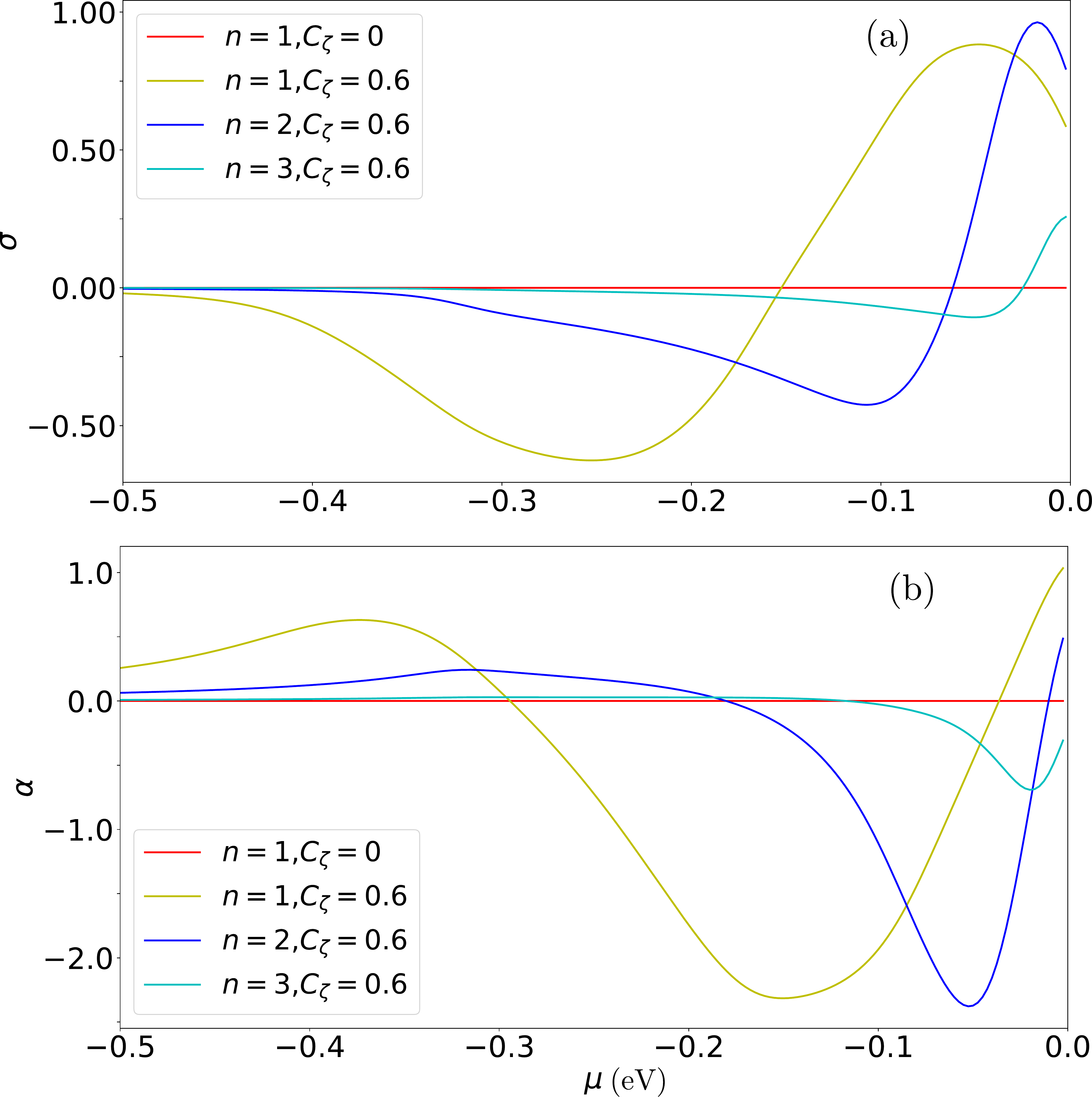}
\caption{ (a) MH (in the unit of $10 e^2/\hbar$) and (b) MN (in the unit of $ ek_{B}/\hbar$) conductivities are shown for tilted and untilted Weyl nodes with $\zeta=+1$. We consider  $v=1$ eV$\cdot$\AA, $\Delta U=0.01$ eV, and $k_{B}T=0.001$ eV. We observe that the  untilted Weyl node with $C_{+}=0$ (tilted Weyl node with $C_{+}=0.6$) results in null (substantial) Magnus responses. We note that MH and MN conductivities are exactly opposite at two opposite Weyl nodes with $\zeta=+1$ and $-1$ owing to the anti-symmetric nature of BC (Eq.~(\ref{eq_bcl})).}
\label{fig:WT}
\end{figure}


The $x$-component of the quasi-particle velocity  associated with the multi-Weyl node is given by
\be
v^{WSM}_{x}({\bm k},\zeta)=  \frac{ k_x n \alpha_n^2 k_{\bot}^{2n-2}}{\epsilon_{\bm k}}.
\label{eq_vel}
\ee
A close inspection suggests that $x$-component of the velocity is also independent of the tilt parameter $C_{\zeta}$ (chirality $\zeta$) like (unlike) the BC. Therefore, the effect of the tilt is only incorporated by the Fermi surface 
properties. The Fermi surface for $n = 1$ case is circular in shape that gets elongated along $k_x$ direction with
increasing $n$. In the $k_x-k_z$
plane, Fermi surface does not close that is in accordance
with the non-point like nature of Fermi surface for tilted WSM. With increasing topological charge  $n$, Fermi surface gets flattened. Therefore, tilt and non-linear dispersion imprint their signature in the transport through Fermi surface properties.

The numerically computed MH responses for  single, double and triple Weyl node as a function of $\mu$ are shown in Fig. \ref{fig:WT}. Interestingly, we find that both MH and MN 
conductivities vanish identically for untilted Weyl node. This is due to the fact that, for a given untilted Weyl node, the positive and negative BC for the activated momentum modes over the Fermi surface are equal, which results in a complete cancellation. This situation remarkably changes in the presence of tilted Weyl node. In this case, the positive and negative BC for the activated momentum modes over the Fermi surface are unequal, and therefore, do not cancel each other completely.

\begin{figure}[h!]
\includegraphics[width=\columnwidth]{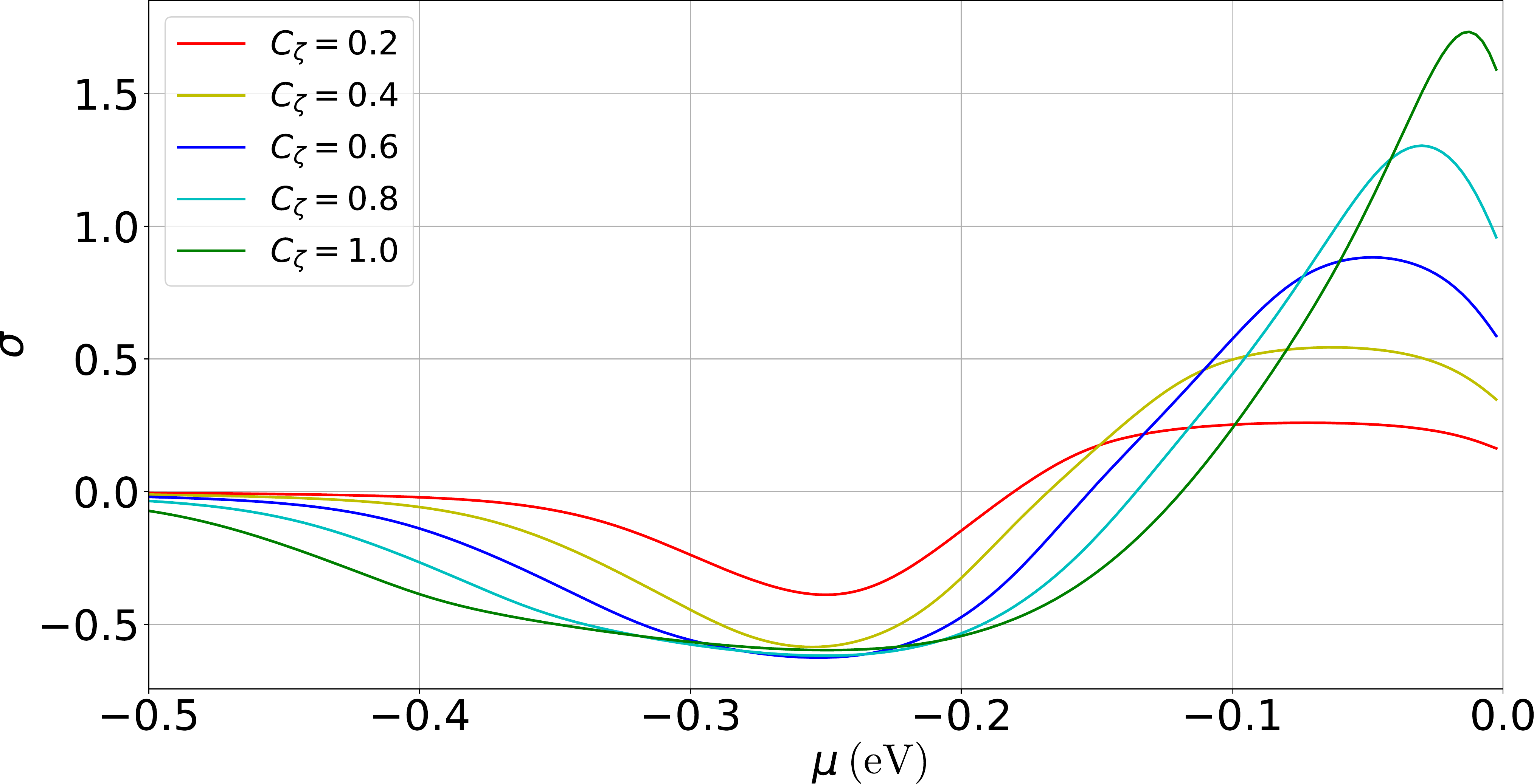}
\caption{(a) MH (in the unit of $10 e^2/\hbar$) conductivity is shown for different strengths of the tilting parameter ($C_{+}$) for fixed $n=1$. Gradual increase in the response is observed with increasing tilt strength $C_{+}$. All other parameters are kept same as mentioned in Fig. \ref{fig:WT}.}
\label{fig:tilt}
\end{figure}

The MH conductivity of a tilted WSM, considering the contribution from two opposite chirality nodes, is given by $\sigma=\sum_{\zeta} G(\mu)  \zeta C_{\zeta}$, where $G(\mu)$ is  $\mu$ dependent part of transport  coefficient 
associated with individual Weyl node. For  a pair of Weyl nodes at same energy $E_0$ such that $G(\mu=E_0)=G_0$, the  MHE is only finite when  relative sign of the tilt parameter between them  
is opposite, referring to the chiral tilt configuration ($C_+=-C_-$). On the other hand, the MHE vanishes in the absence of tilt ($C_+=C_-=0$), and even in the presence of achiral tilt ($C_+=C_-$) of the Weyl node. In other words, MH responses from opposite Weyl node add up (cancel each other) leading to a node integrated (polarized) Magnus  (Magnus valley) response in presence of 
chiral (achiral) tilt.  Two  Weyl nodes of 
opposite chirality, residing at two different energies $E_{+}$ and $E_{-}$, can in principle lead to Magnus valley Hall effect while Weyl nodes exhibit  achiral tilt
such that $G(\mu=E_+) \ne G(\mu=E_-)$. 
The MNE follows the same behavior as MHE.
This is because the sign of BC is opposite whereas the $v_x^{WSM}$ has same sign for two different nodes of opposite chirality.   
 To shed more light into the tilt mediated MH response, we show the systematic growth of MH conductivity while increasing the tilt strength
in Fig. \ref{fig:tilt}. 

Therefore, the MH and MN conductivities 
can become useful probes in distinguishing tilted WSM from an
untilted WSM in experiments. We also notice that the responses for $n=1$ single WSM is found to be most prominent as compared to the WSMs with $n>1$.  This can be explained as the  BC reduces its value for the activated momentum modes over 
the Fermi surface. Moreover,  the MH responses  decrease as the Fermi surface becomes more flattened for mWSMs.  Another interesting feature, coming out from Eq.~(\ref{eq_bcl}) and (\ref{eq_vel}), is that MH responses from two opposite nodes with chiral (achiral) tilt simply add up (cancel each other) leading to a finite node integrated (polarized) transport coefficients 
similar to the Magnus (Magnus valley) responses for
ML graphene in presence (absence)  of warping.  From low energy model, it can be shown that the  node integrated MH responses are proportional to that of a single tilted Weyl node.
All these findings together refer to very interesting Magnus transport properties of WSM in general.

\textcolor{black}{It is pertinent to discuss the anomalous Hall effect that will present alongside with the MHE for the TRS broken WSM as far as the first order responses are concerned. 
The anomalous Hall conductivity  for type-I mWSM is found to be 
$\sigma' \propto n\Delta k$, where  $\Delta k$ is the separation between two Weyl nodes in momentum space of opposite chirality \cite{burkov14,Nag20}. On the other hand, the MHE in type-I mWSM ($|C_{\zeta}/v|\ll 1$) for a given chmeical potential can be analytically found as $\sigma \propto n\frac{\Delta U}{v}$. It is clear from the above expressions that the intrinsic AHE increases  with increasing the $\bm k$-space separation between Weyl nodes while remains insensitive to the tilt parameter. By contrast, the magnitude of MH conductivity increases with increasing the built-in electric field. Using the above expressions, one can in general find $\frac{\sigma}{\sigma'}\propto \frac{\Delta U}{v \Delta k}$ (in an arbitrary unit) for type-I WSM. Since this comparison is based on the low-energy model of mWSM, one should consider the lattice model to make a correct estimate of $\frac{\sigma}{\sigma'}$. In addition, we would like to point out that the intrinsic linear AHE can in principle be found to be quantized  while MHE is not expected to exhibit quantized response \cite{simin20}. }

\section{Conclusions}
\label{cons}
In conclusion, we first investigate the effect of strain and warping on MHE and MNE in ballistic regime for 2D topological systems such as  ML, BL graphene and surface states of TIs. We find that in strained ML graphene system without warping, the total Magnus responses are zero after summing over the valleys because the contribution from the each valley cancels with each other. One instead obtains Magnus valley Hall and Magnus valley Nernst effects from the valley polarized contribution. 
Interestingly, we find that the warping leads to
finite total Magnus responses as BC contributions from each valley to Magnus responses are unequal and do not cancel with each other for the  asymmetric nature of 
Fermi surface shape. The magnitude of the total Magnus responses is found to increase
with increasing the strength of warping parameter. For BL graphene, strain enhances asymmetry between the valley polarized contribution resulting in distinct transport signatures for positive and negative strain, while the effect of warping remains minimal.  
 In the case of surface states of TI, we find that the magnitude of both MH and MN conductivities enhance with increasing the hexagonal warping strength.

Going beyond 2D, we study Magnus responses in ballistic regime for 3D Weyl semimetals using low-energy model to probe the effect of tilt and anisotropic nonlinear dispersion. In particular, we find that the MHE is identically zero for each Weyl node without tilt, whereas for tilted WSMs, Magnus responses coming from the nodes acquire finite values. 
Notably, MH responses from opposite Weyl node add up (cancel each other) leading to a node integrated (polarized) Magnus  (Magnus valley) response in presence of 
chiral (achiral) tilt.
The magnitude of both MH and MN conductivities increase with increasing the tilt parameter of the Weyl nodes. Moreover, with increasing the topological charge associated with Weyl node, the Magnus responses get suppressed.
This key feature can be a useful probe in distinguishing untilted (type-I), tilted (type-I or type-II) WSM and the non-linearity in the dispersion through experiments. Moreover, from the application point of view, such  MH responses can  
pave the way for a new generation of current rectification devices \btext{ where the alternating-current signal is converted into a direct-current signal. This is due to the fact that linear Hall effects, 
the transverse MH voltage (say in $y$-direction) is developed due to the Magnus velocity of the carriers having positive longitudinal velocity ($v_x>0$) only}.

\btext{Related to the experimental realization of MH response in 3D, we would like to first point out that the  bulk quantum Hall effect  has been realized earlier in quasi-2D systems \cite{Cao2012,Masuda2016,Uchida2017}. Recently, 3D systems, such as ZrTe$_5$, HfTe$_5$, and Cd$_3$As$_2$ have been shown to exhibit quantum Hall effect in experiments \cite{Zhang2018,Liang2018,Tang2019,Galeski2020}. Moreover, non-linear Hall effect has already been experimentally observed in bilayer non-magnetic quantum material WTe$_2$ \cite{Ma2018},  in few layer of WTe$_2$ \cite{Kang2019}, and in type-II WSM at room temperature \cite{Kumar2021}. In light of the above experiments, it is in principle possible to extend the non-linear Hall effect setup, fabricated in 2D, to 3D platforms, where MHE can be experimentally observed. For example, 2D systems can be stacked together in order to form quasi-2D / 3D structure over which multi-terminal Hall measurements can be performed. In the case of MHE, the choice of suitable gate potentials, causing the built-in electric field become very important to generate the appropriate transverse Hall voltage}.

In contrast to the linearized model we use in this work, a real mWSM may contain Weyl nodes with different tilt with respect to one another as well as number of pair of nodes can be greater than one. One of the interesting extensions of this work would be to implement the MH responses calculation on a real mWSM material using DFT or at least on a lattice model in order to directly compare with the experiments. For this purpose, one can perform a four terminal Landauer-B{\"u}ttiker conductance calculation \cite{Buttiker86} on a lattice. Following our theoretical analysis on mWSMs, we expect MH responses to be negligible for materials like NbAs, TaAs, which have symmetric untilted Weyl cones. On the other hand, type-II Weyl materials (MoTe$_2$,WTe$_2$) can show substantial MH responses. Moreover, investigating MH responses in twisted BL graphene would be an interesting direction which we leave for future study.

\begin{acknowledgements} S.N. acknowledges the National Science Foundation Grant No. DMR-1853048. S.K.D would like to thank Ulrike Nitzsche for technical assistance.

 \end{acknowledgements}

\bibliography{Magnus_Graphene}

\end{document}